\documentclass[fleqn,usenatbib]{mnras}

\usepackage{newtxtext,newtxmath}

\usepackage[T1]{fontenc}

\DeclareRobustCommand{\VAN}[3]{#2}
\let\VANthebibliography\thebibliography
\def\thebibliography{\DeclareRobustCommand{\VAN}[3]{##3}\VANthebibliography}

\usepackage{graphicx}	
\usepackage{amsmath}	
\usepackage{xcolor}	
\usepackage{caption}
\usepackage{subcaption}
\usepackage{xspace}

\renewcommand{\sun}{\ensuremath{\odot}}
\newcommand{\msun}{\ensuremath{{\mathrm{M}}_{\odot}}\xspace}

 \title[3D radiative transfer kilonova modelling]{3D radiative transfer kilonova modelling for binary neutron star merger simulations}

\author[C. E. Collins et al.]{Christine E. Collins,$^{1}$\thanks{E-mail: c.collins@gsi.de}
Andreas Bauswein,$^{1}$
Stuart A. Sim,$^{2}$
Vimal Vijayan,$^{1,3}$
Gabriel Mart\'inez-Pinedo,$^{1,4}$ \newauthor
Oliver Just$^{1,5}$,
Luke J. Shingles$^{1}$
and Markus Kromer$^{6}$
    \\
    $^{1}$GSI Helmholtzzentrum f\"{u}r Schwerionenforschung, Planckstraße 1, 64291 Darmstadt, Germany\\
    $^{2}$Astrophysics Research Center, School of Mathematics and Physics, Queen's
    University Belfast, Belfast BT7 1NN, Northern Ireland, UK\\
    $^{3}$Department of Physics and Astronomy, Ruprecht-Karls-Universit\"at Heidelberg, Im Neuenheimer feld 226, 69120 Heidelberg, Germany\\
    $^{4}$Institut f{\"u}r Kernphysik (Theoriezentrum),
    Fachbereich Physik, Technische Universit{\"a}t
    Darmstadt, Schlossgartenstra{\ss}e 2, 64289 Darmstadt, Germany \\
    ${^5}$Astrophysical Big Bang Laboratory, RIKEN Cluster for Pioneering Research, 2-1 Hirosawa, Wako, Saitama 351-0198, Japan \\
    $^{6}$Heidelberger Institut f\"ur Theoretische Studien (HITS), Schloss-Wolfsbrunnenweg 35, 69118 Heidelberg
}

 \date{Accepted XXX. Received YYY; in original form ZZZ}

 \pubyear{2022}

\begin{document}
\label{firstpage}
\pagerange{\pageref{firstpage}--\pageref{lastpage}}
\maketitle

\begin{abstract}
The detection of GW170817 and the accompanying electromagnetic counterpart, AT2017gfo, have provided an important set of  observational constraints for theoretical models of neutron star mergers, nucleosynthesis, and radiative transfer for kilonovae.
We apply the 3D Monte Carlo radiative transfer code ARTIS to produce synthetic light curves of the dynamical ejecta
from a neutron star merger, which has been modelled with 3D smooth-particle hydrodynamics (SPH) and included neutrino interactions.
Nucleosynthesis calculations provide the energy released from radioactive decays of r-process nuclei, and radiation transport is performed using grey opacities given as functions of the electron fraction.
We present line-of-sight dependent bolometric light curves, and
find the emission along polar lines of sight to be up to a factor of $\sim 2$ brighter than along equatorial lines of sight.
Instead of a distinct emission peak, our bolometric light curve exhibits a monotonic decline, characterised by a shoulder at the time when the bulk ejecta becomes optically thin.
We show approximate band light curves based on radiation temperatures
and compare these to the observations of AT2017gfo.
We find that the rapidly declining temperatures lead to a blue to red colour evolution similar to that shown by AT2017gfo.
We also investigate the impact of an additional, spherically symmetric secular ejecta component, and we find that the early light curve remains nearly unaffected, while after about $1\,$ day the emission is strongly enhanced and dominated by the secular ejecta, leading to the shift of the shoulder from $\sim$1-2 to 6-10 days.
\end{abstract}

\begin{keywords}
Radiative transfer -- (Transients:) neutron star mergers -- Methods: numerical
\end{keywords}



\section{Introduction}

The detection of GW170817 \citep{abbott2017a} and its optical counterpart
AT2017gfo \citep[e.g.][see \citealt{villar2017a} and references therein]{smartt2017a} confirmed the prediction that a kilonova would accompany the 
merging of binary neutron stars (BNS) (\citealt{li1998a, metzger2010a}, see e.g. \citealt{metzger2019a} for a recent review).
The observations of a power-law luminosity decline consistent with r-process material and a sufficient event rate are consistent with binary neutron star mergers being the dominant site of
r-process element production in the Universe \citep{kasen2017a, drout2017a},
which had previously been proposed by some authors on theoretical grounds \citep{lattimer1976a, eichler1989a}.

Simulations of BNS mergers predict complex ejecta structures. The merger ejecta is comprised of material expelled on dynamical
timescales of tens of milliseconds, which is followed by secular ejecta expelled on timescales of seconds.
The dynamical ejecta typically have mass of
$10^{-4}$ -- $10^{-2}$ \msun, and high ejecta velocities,
typically 0.2 -- 0.3c  \citep[e.g.][]{bauswein2013a, hotokezaka2013a, tanaka2013a, kruger2020a, radice2018a,ardevol2019a}.
The secular ejecta, i.e. the material unbound on longer time scales, are typically more massive
($10^{-2}$ -- $10^{-1}$ \msun) and ejected with lower velocities
of $\sim$0.1c \citep[e.g.][]{fernandez2013a, perego2014a, just2015a, fujibayashi2018a, siegel2018a}.

The ejecta are predicted to have low lanthanide fraction material
in the polar directions, where there is a lower neutron abundance,
corresponding to electron fractions of $Y_{e} \gtrsim 0.3$, which would lead to low opacities and therefore `blue' colours in the predicted kilonova.
Higher lanthanide fractions are expected in the equatorial directions,
where there is a higher neutron abundance ($Y_{e}$ $\lesssim 0.3$)
which would lead to higher opacities and `red' colours in the predicted kilonova \citep[e.g.][]{metzger2014a, sekiguchi2015a, just2015a, perego2014a, foucart2020a, radice2022a}.

The observations of the kilonova AT2017gfo initially showed blue colours
with the spectra peaking in the UV/blue,
which rapidly evolved to redder colours with spectra peaking in the near-infrared \citep[see e.g.][]{smartt2017a, villar2017a}.
It has been suggested that to explain this rapid colour evolution,
the outermost layers of the merger ejecta must be composed of 
low opacity, high $Y_{e}$ material (which would lead to low lanthanide
fraction material), while the inner ejecta layers require higher opacity, 
low $Y_{e}$ material (high lanthanide fraction) to reproduce the red colours \citep[e.g.][based on the analytical two compnent model introduced by \citealt{metzger2017a}]{cowperthwaite2017a, villar2017a}.

Initial progress in this field has relied on the use of
analytic ejecta models that are described by e.g., power law density structures
or idealised geometries \citep[e.g.,][]{metzger2010a, barnes2013a, watson2019, banerjee2020a, even2020a, domoto2021a, heinzel2021a, korobkin2021a, wollaeger2021a, pognan2022a}.
Kilonova studies based on parameterised ejecta configurations avoid the complexities connected to hydrodynamics modelling, while other aspects, such as thermalisation (e.g. \citealt{barnes2016a}) can be studied in more detail.
Ultimately, however, a reliable interpretation of future multi-messenger observations of BNS will require kilonova models to be based on self-consistent simulations of the merger and its ejecta.

Of the studies that have considered simulated ejecta from BNS mergers,
these have mostly carried out radiative transfer simulations in
1D 
\citep[][]{curtis2021a, gillanders2022a, wu2022a}
and 2D \citep[][]{kasen2015a, kawaguchi2021a, just2022a, kawaguchi2022a, klion2022a},
although we note some of these were based on 3D merger simulations \citep[e.g.][]{kawaguchi2021a, just2022a}.
3D radiative transfer calculations have been carried out by \citet{tanaka2013a},
but for axisymmetric merger data, and by \citet{neuweiler2022a}.
\citet{bulla2021a} {and \citet{darbha2021a}} have carried out 3D radiative transfer simulations
for 3D merger {simulations} of a BH-NS system, and {\citet{nativi2021a} carried out 3D radiative transfer simulations} for a 3D simulation of neutrino-driven winds from a neutron star merger remnant.

In this paper, we carry out radiative transfer simulations for
merger ejecta extracted directly from a BNS
simulation.
Our hydrodynamic simulation does not only self-consistently provide the density distribution of the dynamical ejecta, but also -- thanks to its sophisticated neutrino treatment -- the electron fraction, which determines the r-process composition and therefore the energy powering the kilonova,
as well as the opacities.

\section{Methods}

\subsection{Hydrodynamical Model}
\label{sec:model}

The model we consider in this paper is of the dynamical ejecta
produced from the merger of binary neutron stars.
The merger simulation was carried out using 
a 3D general relativistic smooth-particle
hydrodynamics (SPH) code \citep{oechslin2002a,bauswein2013a},
and an advanced neutrino leakage treatment, ILEAS \citep[Improved leakage-equilibration-absorption scheme][]{ardevol2019a}.
As a first step, 
we consider equal mass neutron stars, each of 1.35 M$_\sun$
(gravitational mass for an orbit with infinite separation),
and each with 150000 SPH particles, using the SFHo equation of state
\citep{steiner2013a}. Hence, the chosen setup represents a system well compatible with GW170817 considering its mass and equation of state.
The stars were placed on an orbit with a separation of $\sim$38~km and we assumed no intrinsic spin of the stars.
The stellar matter was initially cold and in neutrinoless beta-equilibrium.
The simulation was evolved until 0.02 seconds after the time when both stars first touched. The calculation thus only covers the early dynamical mass ejection, while considerable amounts of ejecta may still be produced on longer time scales. The system did not collapse to a black hole within the timescale of the simulation.
For the given mass and equation of state a delayed gravitational collapse may be expected.

\subsubsection{Mapping SPH particles to a grid}
\label{sec:mappingparticles}

The SPH particles from the merger simulation
must be mapped onto a Cartesian grid for the radiative transfer simulation.
We first propagate the particles based on the velocities they had
at the end of the SPH simulation for a further 0.5 seconds.
After this time we assume homologous expansion
and map the particle positions onto a 128$^3$ Cartesian grid spanned in velocity space.
We set the maximum absolute velocity of the velocity grid (i.e. the grid boundary)
equal to 0.5c (along each axis, noting that the corners of the Cartesian grid extend to velocities up to 0.87c).
The bulk of the ejecta mass is well below this velocity.
Few SPH particles are at velocities higher than 0.5c, corresponding to 5\% of the total ejecta mass,
and we do not expect
these high velocity particles at very low densities to
have a strong influence on the predicted light curves at the times we consider in our simulation.
We map the density and $Y_e$ of the SPH particles to the Cartesian grid. The mapping of a quantity $A$ is performed via
\begin{equation}
    A(r)=\frac{1}{N}\sum_a A_a \frac{m_a}{\rho^*_a}W(|r-r_a|;h)
\end{equation}
with the sum running over all particles $a$, whose kernel function $W$ is finite at the position $r$. Here $A_a$ is the value of A associated with particle $a$, $m_a$ is the particle mass, $\rho^*_a$ is the conserved rest-mass density, $r_a$ is the position of particle $a$,
$W$ is the SPH kernel function with smoothing length $h$. $N$ is introduced for normalization and is defined by $N=\sum_a \frac{m_a}{\rho^*_a}W(|r-r_a|;h)$. For the density we do not use the normalization but compute $\rho(r)=\sum_a m_a W(|r-r_a|;h)$ (this expression includes an additional approximation, namely equating $\rho$ and $\rho^*$).
For the mapping, we adopt the smoothing length and smoothing kernel from the SPH simulation.
In order to reduce the clumpiness of the outermost ejecta (likely related to the limited particle
resolution used in the underlying SPH simulation) we increase the smoothing length by 50\% for particles
with absolute velocities {$v > 0.23$c}.

\subsubsection{Dynamical ejecta}

\begin{figure*}
\centering

\includegraphics[width=0.75\textwidth]{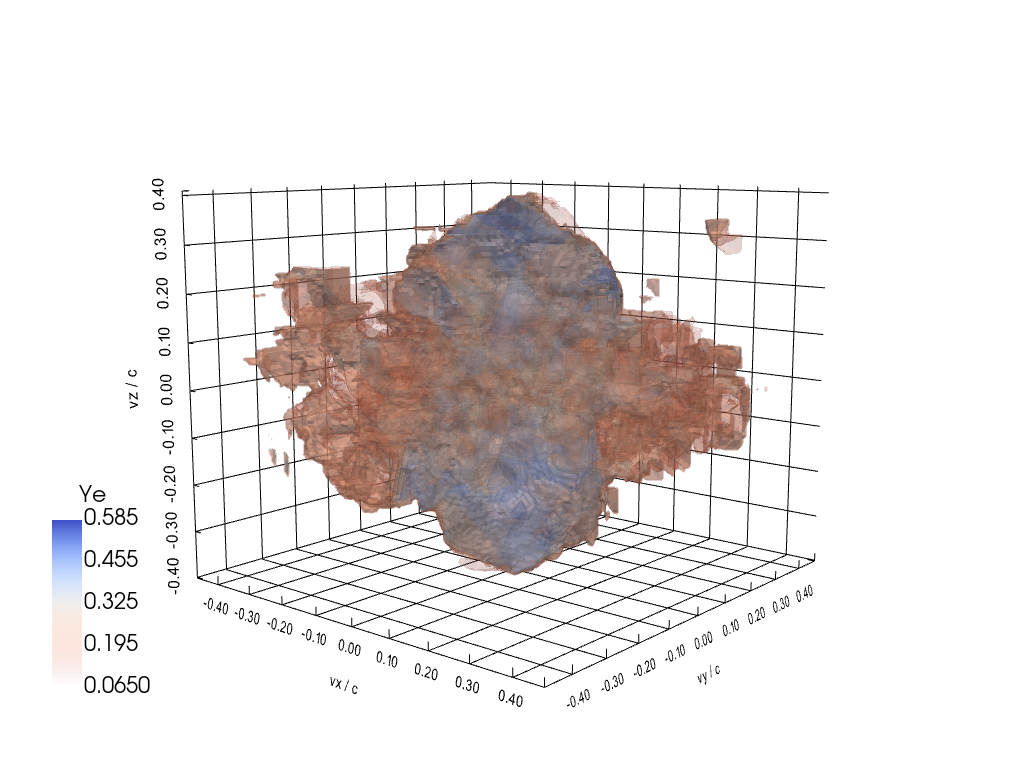}

\caption{3D rendering of the dynamical ejecta mapped to a Cartesian grid.
The v$_{\rm z}$ axis corresponds to the rotation axis (or polar axis) of the system and v$_{\rm z} = 0$ represents the equatorial plane.
For visual purposes, in this figure we do not show grid cells containing density contributions from less than 5 SPH particles. 
The colour scale is set by isosurfaces showing
the average $Y_{e}$ in a grid cell.
Material near the equator generally has a lower $Y_{e}$ than in the polar directions.
}

\label{fig:ejecta_3D_Ye}
\end{figure*}

\begin{figure}
\centering
\includegraphics[width=0.4\textwidth]{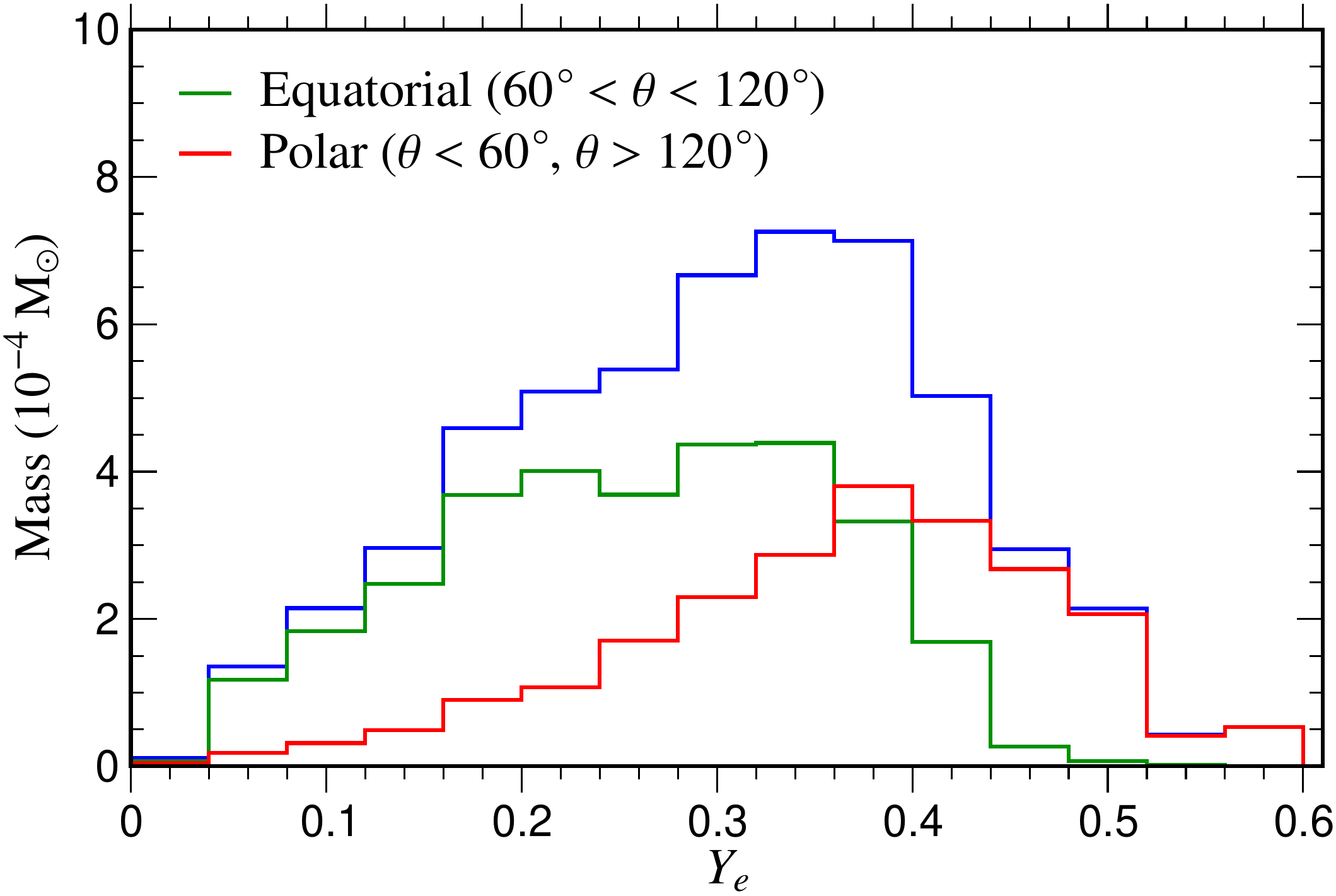}

\caption{Distribution of ejecta mass (of all SPH trajectories) into electron fraction bins ($\Delta$$Y_e=0.04$) (blue),
    and separated into polar (red) and equatorial regions (green).
    The electron fraction near the poles is generally higher than near the equator.
}

\label{fig:Ye-mass}
\end{figure}

\begin{figure}

\includegraphics[width=0.5\textwidth]{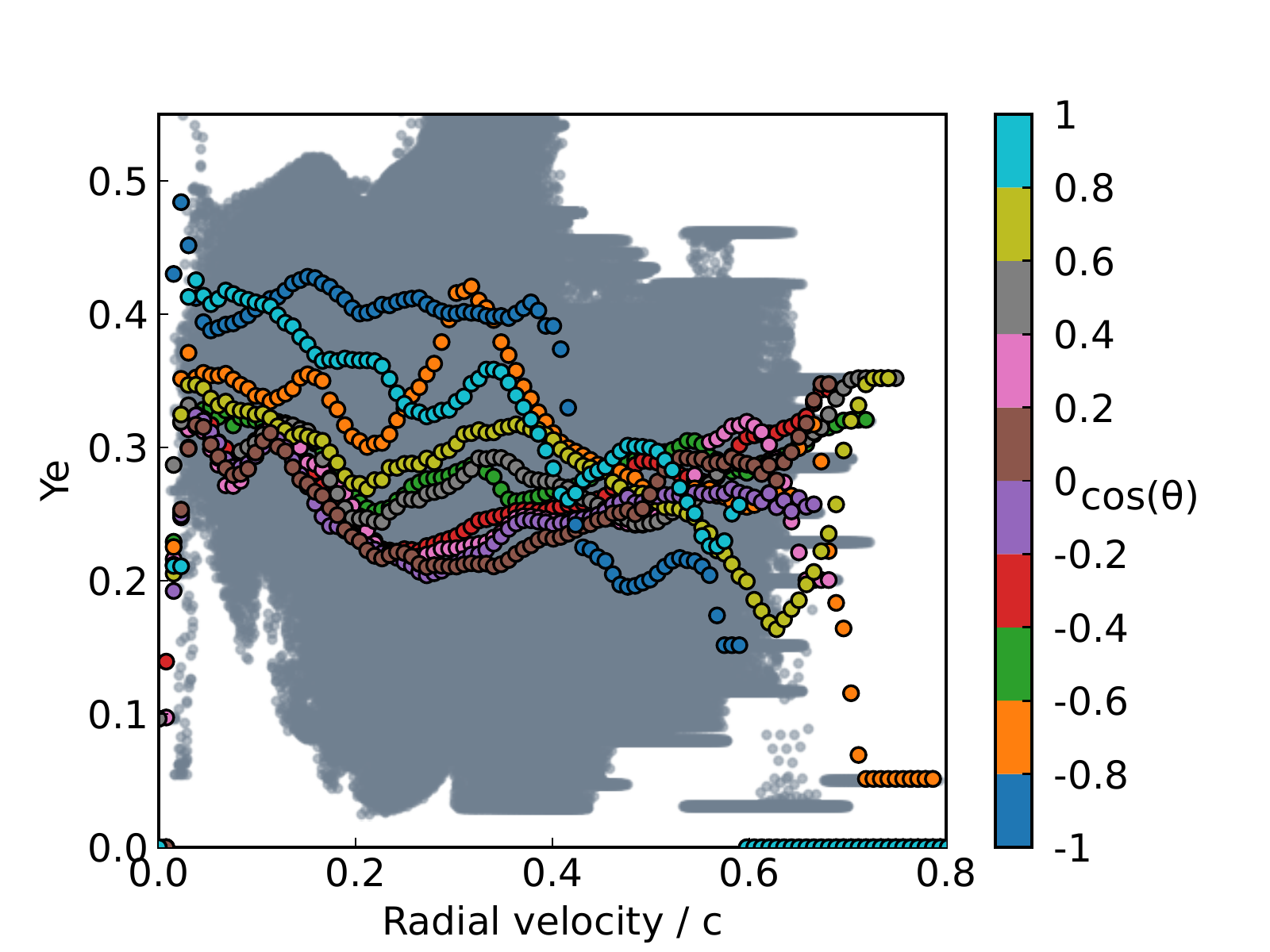}

\caption{Electron fraction ($Y_e$) in each model grid cell on the Cartesian grid, plotted against absolute velocity.
We mark the average $Y_e$ within 10 uniform angle bins to indicate angle dependence.
We note that due to the low resolution of the outer ejecta, at high velocities we find that one SPH particle can influence the $Y_e$ of many grid cells
(shown in this plot by a line at the same $Y_e$ for a range of velocities).
}

\label{fig:Ye-everycell}
\end{figure}

\begin{figure}

\includegraphics[width=0.48\textwidth]{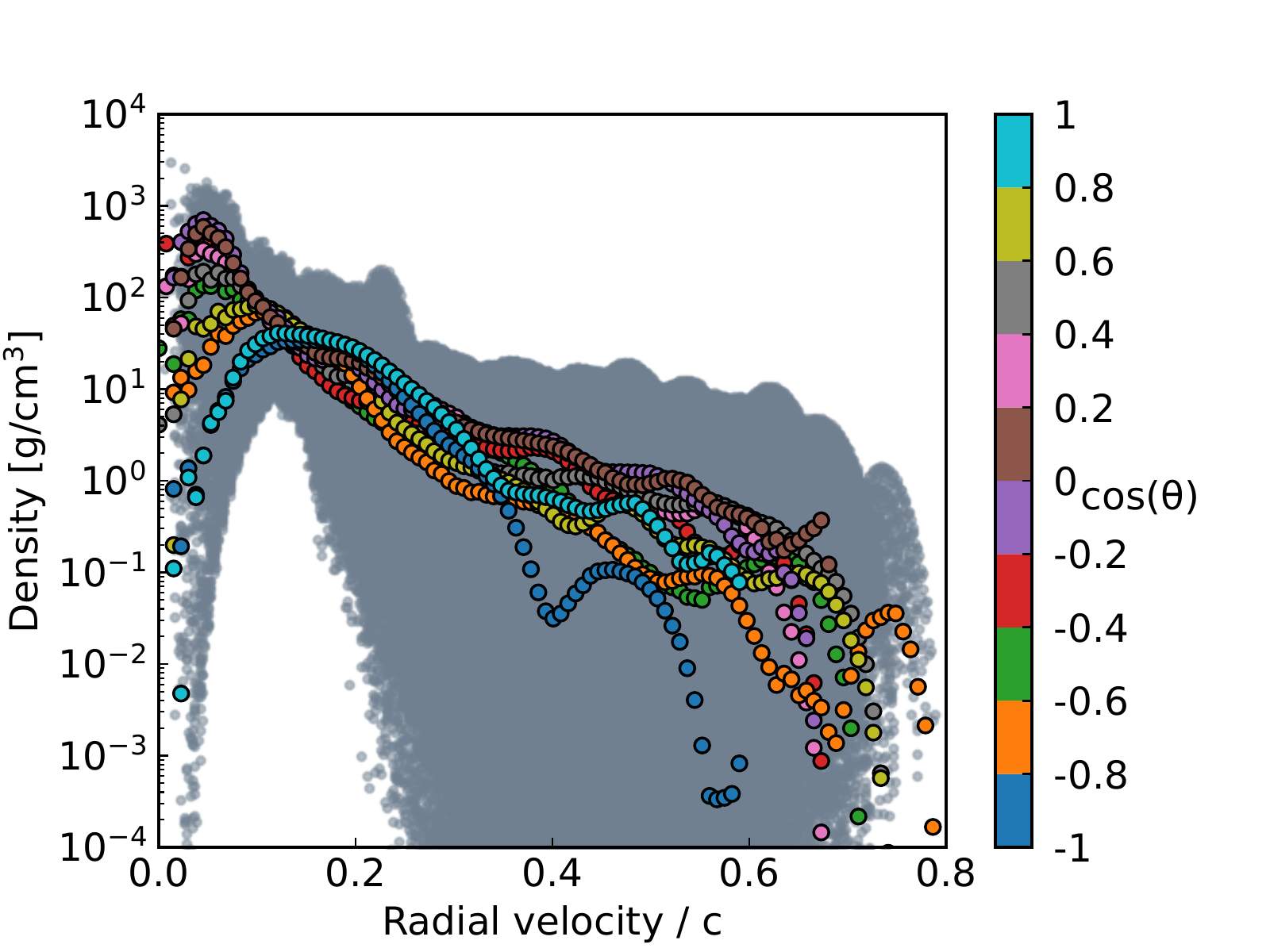}

\caption{Cell mass densities (on the Cartesian grid) versus absolute radial velocity
at 0.52 seconds after the merger, corresponding to the time at which SPH particles are mapped to the
Cartesian grid and homologous expansion is assumed.
The average densities in ten polar angle bins have been plotted in colour to indicate the density range along different lines of sight.
The outer ejecta are less well resolved numerically, leading to scatter in the densities at high velocities.
}

\label{fig:rho-everycell}
\end{figure}
The ejecta is resolved by $\sim$2000 SPH particles, which become unbound during the simulation
(see \citealt{oechslin2002a,bauswein2013a} for ejecta criterion).
The total mass of dynamical ejecta mapped onto the radiative transfer grid was 0.0051 \msun \ (although we note that mass ejection through different mechanisms still continues after the end of the merger simulation).
In Figure~\ref{fig:ejecta_3D_Ye} we show a 3D rendering of
the dynamical ejecta once it has been mapped to the grid, indicating the 3-dimensional structure
produced by the merger simulation.
The colour scale indicates the $Y_{e}$ of the ejecta (where $Y_{e}$ is defined at the end time of the merger simulation).
Generally, lower $Y_{e}$ material is found near the equator,
while higher $Y_{e}$ material is found in the polar directions.
This is demonstrated by Figure~\ref{fig:Ye-mass},
showing the $Y_{e}$ of the trajectories at the end of the SPH simulation.
The trend of decreasing $Y_e$ from the pole to the equator is consistent with previous works,
e.g. \citet{radice2018a, foucart2020a, kullmann2022a}.
However, we also find that lower $Y_{e}$ material is mixed with the
higher $Y_{e}$ material in the poles (i.e. the polar directions are not
composed solely of high $Y_{e}$ material).
A similar effect was reported by \citet{just2022a}.
As a result we expect lower opacities in the polar directions, due to the lower lanthanide fraction expected to be synthesised for the lower $Y_{e}$ material.
The $Y_e$ mapped to the Cartesian grid is shown in
Figure~\ref{fig:Ye-everycell}.
To analyse the 3D density structure of the ejecta, we plot the density of each model grid cell
in Figure~\ref{fig:rho-everycell}.
To indicate the angular dependence, we divide the ejecta into 10 uniform solid-angle bins in the polar direction
relative to the positive z-axis and plot the average density within
each angle bin.
Due to the limited resolution of the underlying SPH simulation, the outermost velocities
may suffer from purely numerical particle noise.
The impact of this particle noise will need to be assessed in future studies using a higher resolution.
While we would expect there to be some difference in the densities in each polar
direction due to statistical fluctuations,
we can not rule out that the relatively large difference
in density between the poles in this model could be numerical.

\subsection{Nucleosynthesis Calculations}

To obtain the energy released by the merger simulation,
time dependent nucleosynthesis calculations are carried
out for each SPH particle trajectory.
The simulation provided thermodynamical histories for all the SPH
particles only up to $t_0 \sim 20$~ms after merger. For later times,
we extrapolate the density evolution assuming homologous expansion,
i.e.\ $\rho(t) r(t)^3 = \rho_0 r_0^3$ and $r(t) = r_0 + v_0 (t-t_0)$
with the subscript $0$ denoting the end of the simulation data. This
corresponds to a density evolution as:

\begin{equation}
  \label{eq:rhot}
  \rho(t) = \rho_0 \left(\frac{\Delta+t_0}{\Delta+t}\right)^3,
\quad \Delta = \frac{r_0}{v_0} - t_0.
\end{equation}
We start the nucleosynthesis calculations at a temperature of $T=10$~GK or the lowest
temperature reached in the SPH simulation if this is higher. Under these
conditions our initial composition is well described by nuclear
statistical equilibrium.
During the nucleosynthesis calculations, the
density is evolved based on the SPH simulation data together with
Equation~(\ref{eq:rhot}). We use the same nuclear reaction network
as in \citet{Mendoza-Temis.Wu.ea:2015} together with the set of
nuclear reactions labelled ``FRDM''. Briefly, it consists of
neutron-capture and photodissociation rates computed within the
statistical model using the FRDM
masses~\citep{Moeller.Nix.ea:1995}. For nuclei with experimentally
unknown $\beta$-decay rates, we use the compilation of~\citet{Moeller.Pfeiffer.Kratz:2003}.
Fission rates~\citep{Panov.Korneev.ea:2010} have been computed based on the
Thomas-Fermi fission barriers of \citet{Myers.Swiatecki:1999}.
Finally, $\alpha$-decay rates
are computed using a Viola-Seaborg formula~\citep{Dong.Ren:2005} for
those nuclei without experimental values.
We use the $Y_e$ value at the end of the SPH simulation to determine the composition at
the beginning of the network calculations. 
{This assumes that Ye is constant between the start of the nucleosynthesis calculations and the end of the SPH simulation data. This is indeed predicted by the simulations as during this phase the expansion timescale is much shorter than the weak interaction timescale.} 

\begin{figure}

\includegraphics[width=0.48\textwidth]{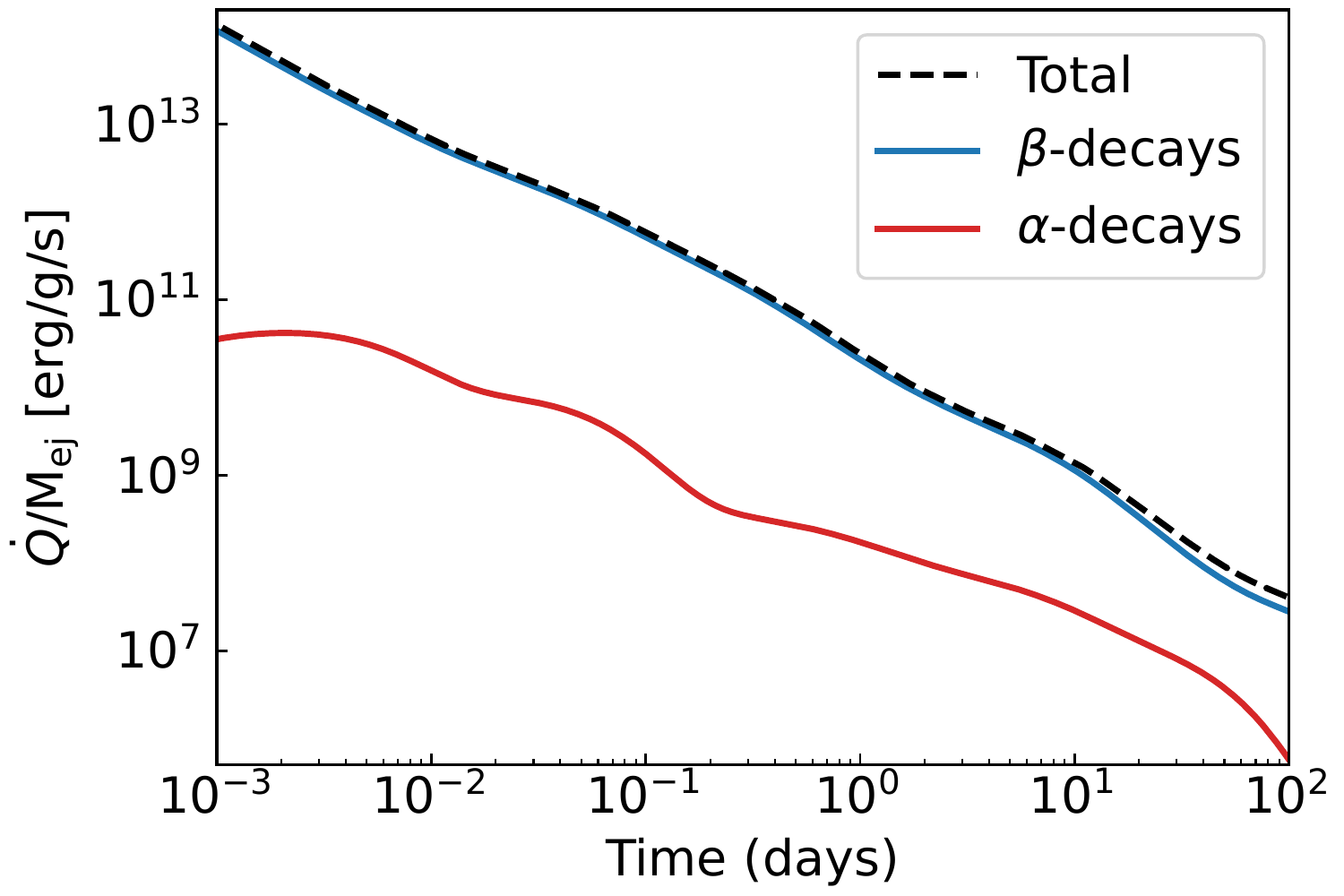}

\caption{Heating rates from nucleosynthesis calculations. The total energy
released (on average for all of the trajectories) is shown by the dashed line.
We also show the total energy from $\beta$-decays and from $\alpha$-particles.
}

\label{fig:trajectoriesQdot}
\end{figure}

In Figure~\ref{fig:trajectoriesQdot}
we show the specific heating rate, $\dot{\mathrm{Q}}$,
averaged over all trajectories.
$\dot{\mathrm{Q}}$ represents the total rate of energy released into the ejecta from nuclear reactions (including the contribution from neutrinos).
Also plotted are the average energies released
from $\beta$-decays and $\alpha$-decays.
Fission also contributes to the total energy released,
although the contribution is orders of magnitudes less than
$\beta$-decays and $\alpha$-decays by $10^{-3}$ seconds,
and is most significant at very early times.
Energy from $\beta$-decays accounts for most of the total
heating rate in the time range considered here.

\subsection{Radiative transfer}
\label{sec:rad_transfer}

We use the time-dependent, multi-dimensional
Monte Carlo radiative
transfer code, \textsc{artis} \citep[][based on the methods of
\citealt{lucy2002a, lucy2003a, lucy2005a}]{sim2007b, kromer2009a}
to predict the kilonova from the neutron star merger
simulation described in Section~\ref{sec:model}.
We propagate $1.15 \times 10^8$ Monte Carlo packets between 0.02 and 120 days after the merger.
We produce viewing-angle dependent light curves by assigning escaping UVOIR-packets into time and directional bins. 
The bin sizes have been chosen to keep Monte Carlo noise at a low level for the observables we present here.
We define 100 uniform solid-angle bins in the polar and azimuthal directions.

\subsubsection{Opacity treatment}

\begin{table}
\centering
\caption{Mass absorption cross sections ($\kappa$) adopted in this study for each electron fraction $Y_{e}$ range. The values are motivated by the Planck-mean opacities calculated by \citet[][table 1]{tanaka2020a}. These calculations assumed a temperature range of
5000 - 10000 K
and a density of $\rho = 1 \times 10^{-13}$ g cm$^{-3}$.
*The lowest $Y_{e}$ opacity is underestimated due to
lack of complete atomic data for actinides \citep{tanaka2020a}.
}

\begin{tabular}{lc}
\hline
$Y_{e}$                  & $\kappa$ \\
                    & cm$^2$ g$^{-1}$ \\ \hline
$Y_{e}$ $\le 0.1$        & 19.5*   \\
$0.1 < $ $Y_{e}$ $\le 0.15$ & 32.2    \\
$0.15 < $ $Y_{e}$ $\le 0.2$           & 22.3    \\
$0.2 < $ $Y_{e}$ $\le 0.25$          & 5.60     \\
$0.25 < $ $Y_{e}$ $\le 0.3$           & 5.36    \\
$0.3 < $ $Y_{e}$ $\le 0.35$          & 3.30     \\
$Y_{e}$ > 0.35                   & 0.96   \\ \hline
\end{tabular}
\label{tab:opacities}
\end{table}

We adopt a temperature independent grey-opacity treatment for the propagation of all radiation
(described by \citealt{sim2007b}).
We use $Y_{e}$ dependent grey absorption cross-sections, based on the
Planck mean opacities listed in table 1 of \citet{tanaka2020a}.
These mean opacities were calculated for temperatures of 5000 - 10000 K
(with $\rho = 1 \times 10^{-13}$ g cm$^{-3}$ at a time of 1 day).
Specifically, we use the opacities listed in Table \ref{tab:opacities}
for model cells with an average $Y_{e}$ within the ranges in Table~\ref{tab:opacities}.
We note that the $Y_e$ in each grid cell is based
on the $Y_e$ of the SPH particles at the end of the merger simulation (not the $Y_e$ evolved in the nuclear network calculation).

\subsubsection{Energy from nucleosynthesis calculations}

\begin{figure}

\includegraphics[width=0.48\textwidth]{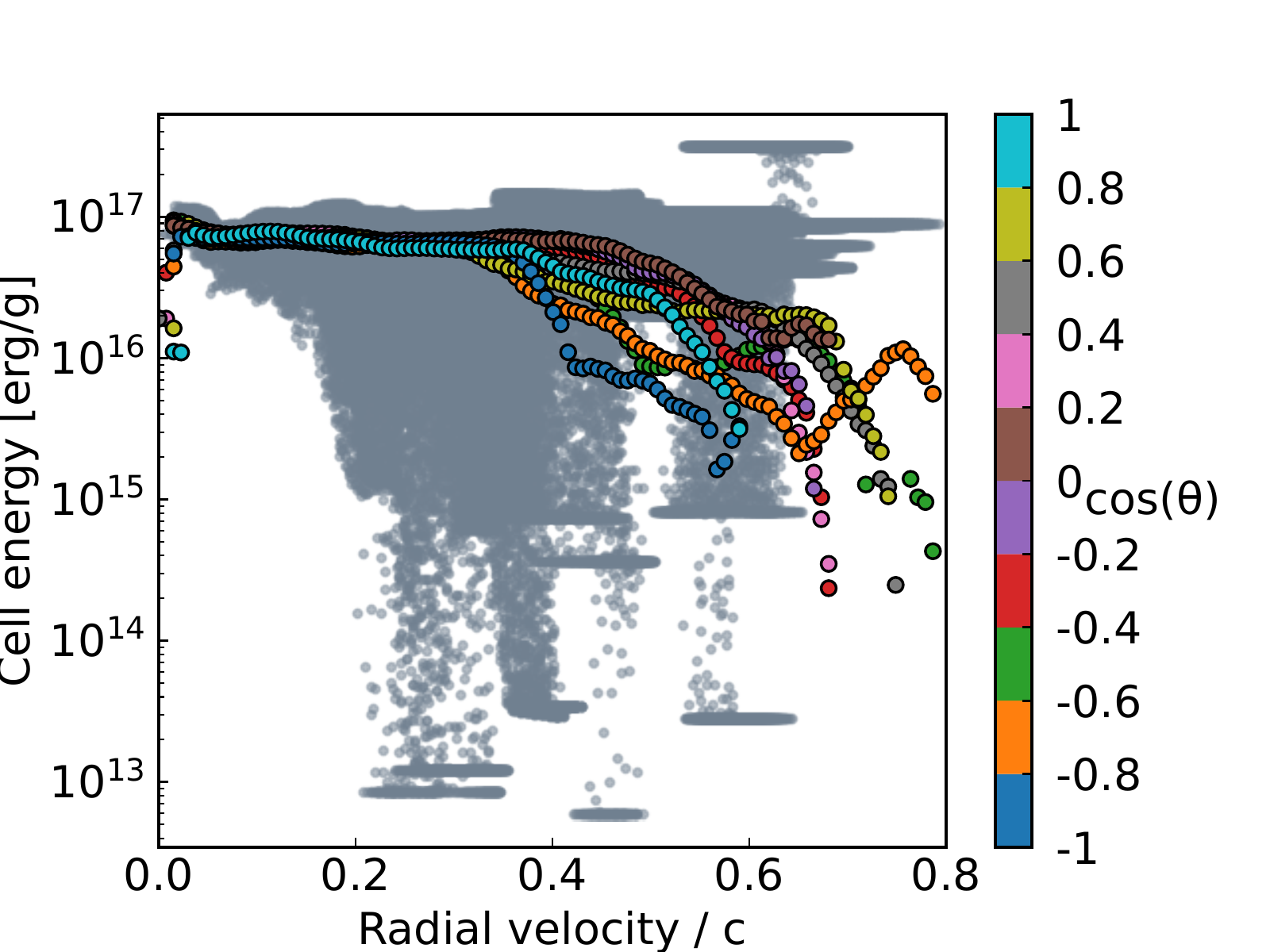}
\includegraphics[width=0.48\textwidth]{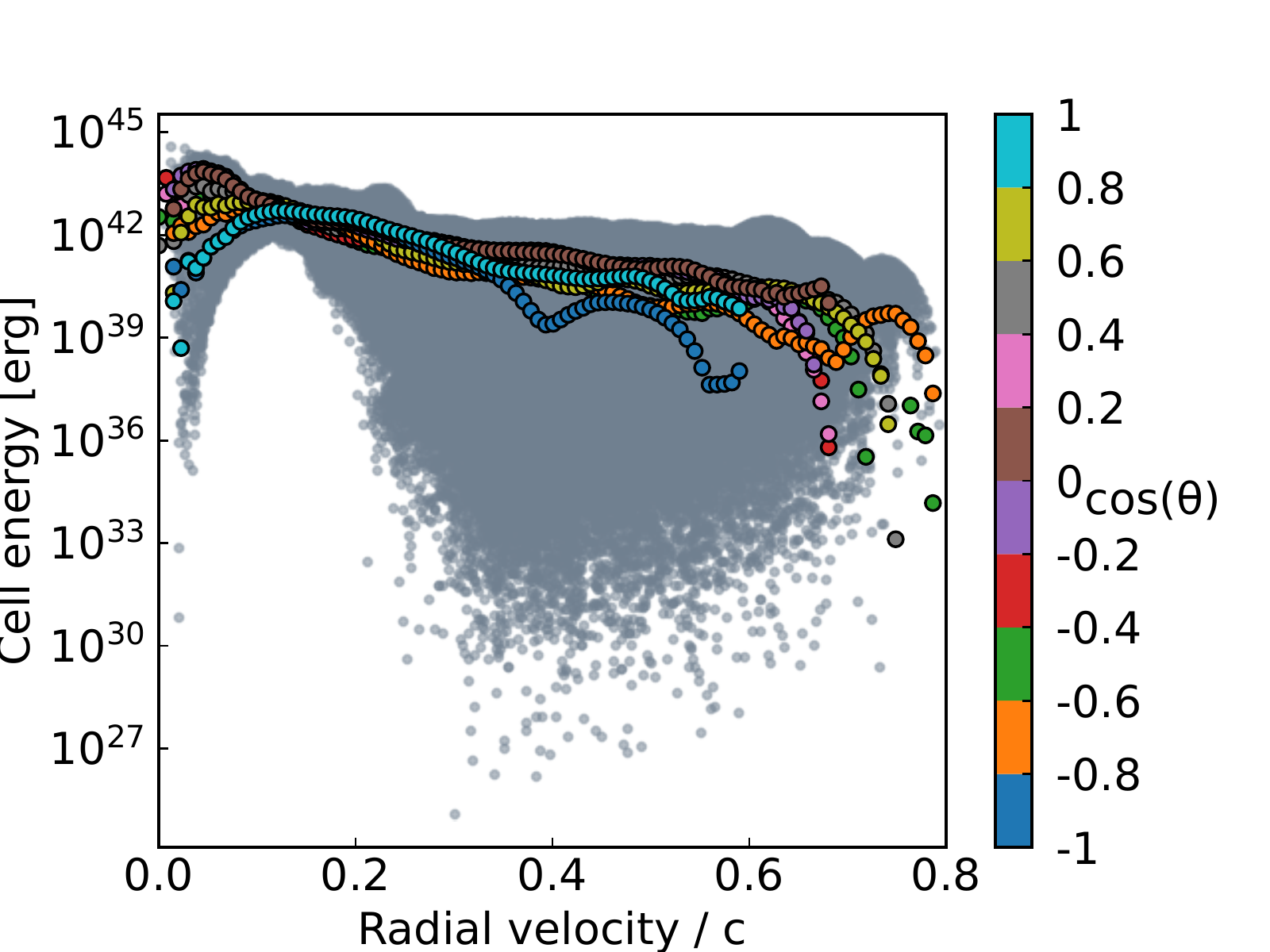}

\caption{Total energy deposited between 0.02 and 120 days versus radial velocity. Top: Deposited energy per unit mass. Bottom: Deposited energy per cubic grid cell (constant velocity extent in x-y-z). Mean values within polar angle bins are indicated in colour.
We note that due to the low resolution of the outer ejecta, at high velocities we find that one SPH particle can influence many grid cells
(shown in the upper plot by a line at the same energy for a range of velocities).
}

\label{fig:qdot}
\end{figure}

In \textsc{artis}, the total energy in the simulation
is defined at the start of the simulation.
We then create Monte Carlo packets of equal energy which
are propagated through the simulation grid.
We define what fraction of the total energy will be deposited in each model grid cell, and place packets according to this distribution.

We define how energy is distributed in the model grid cells
according to the total energy released by the SPH particle
trajectories that contribute to that model grid cell.
The total amount of energy released along a trajectory over time
is mapped to the 3D Cartesian grid using the same 
method as described in Section~\ref{sec:mappingparticles}.
These energies are shown in Figure~\ref{fig:qdot}.
This energy (in erg/g) and the total mass in the model grid cell are used to set the total energy generated in each model 
grid cell, and in the framework of the code, the fraction of energy which will be deposited in the cell
(also shown in Figure~\ref{fig:qdot}).

At the start of the simulation we also define the rate at which energy will be deposited (the time at which a Monte Carlo packet will be placed in a model grid cell).  
We assume that the energy in all model grid cells
is deposited at the same rate, which we take as the
average $\dot{\mathrm{Q}}$ of all unbound SPH particle trajectories
(plotted in Figure~\ref{fig:trajectoriesQdot}).
We consider energy released from 8 seconds after the merger
($\ll$ start time of radiative transfer simulation) until 120 days after the merger.
Energy released before the start of the simulation
is accounted for by placing Monte Carlo packets in the ejecta
which are advected with the homologous flow.
The energy of these packets is reduced to account for
adiabatic losses before the beginning of the simulation.

Since $\beta$-decays are the dominant source
of energy at the times we are considering,
we assume that all of the energy in our simulations
comes from $\beta$-decays.
Following the results of \citet{barnes2016a},
we assume that 35\% of the $\beta$-decay energy emerges
as neutrinos, 20\% as $\beta$-particles and 45\%
as $\gamma$-rays.
The energy from the neutrinos will never thermalise,
so this energy is ignored in the simulations.

\citet{barnes2016a} find that low energy $\beta$-particles effectively
thermalise on kilonova timescales,
and that even slightly tangled magnetic fields are effective at
trapping high energy $\beta$-particles.
We assume that all of the energy from the $\beta$-particles
will thermalise locally and instantaneously.
In reality, the $\beta$-particles will continuously
deposit energy over some distance through Coulomb interactions
with thermal electrons, by ionising or exciting bound atomic
electrons or by Bremsstrahlung emission.
The thermalisation of $\beta$-particles
is dependent on the magnetic field structure
in the ejecta, which we neglect.

\begin{figure}

  \includegraphics[width=0.45\textwidth]{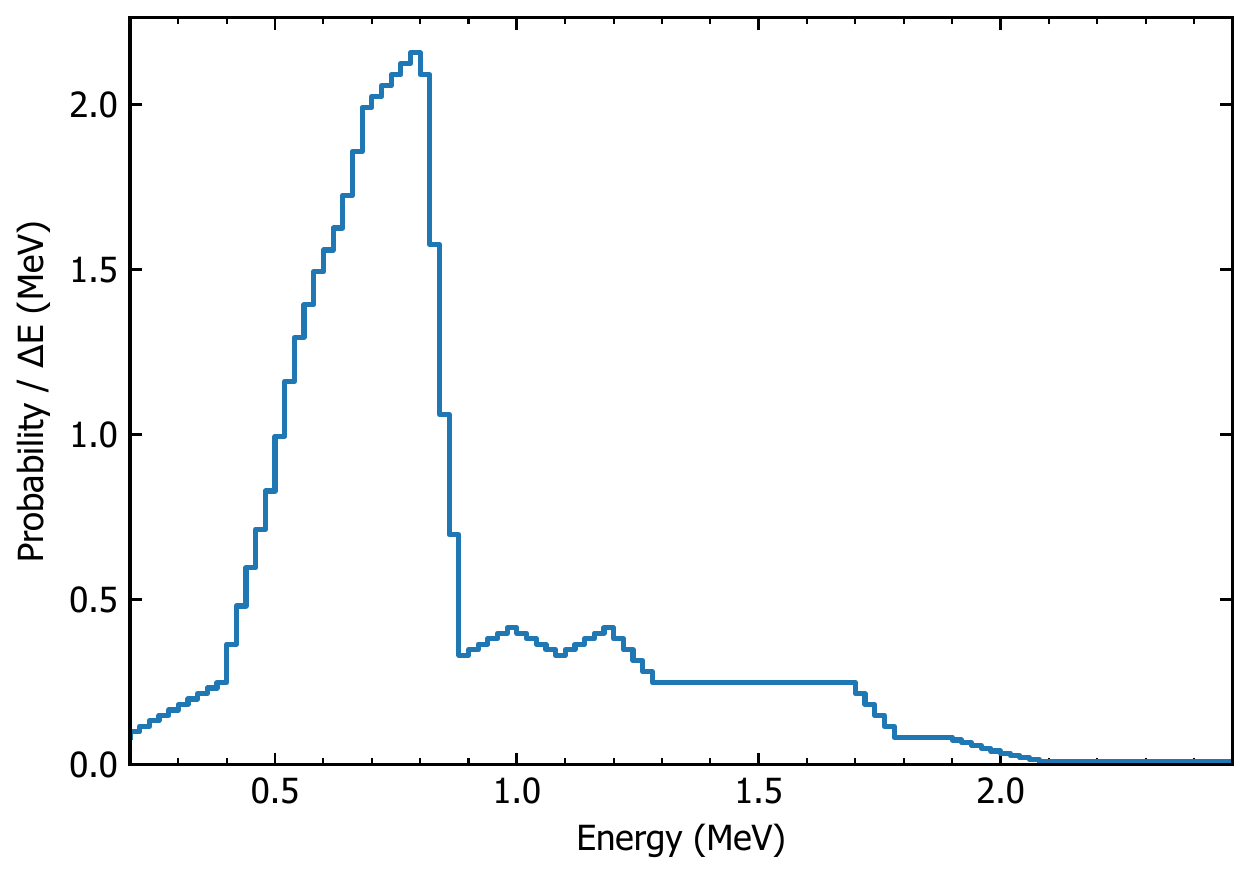}

\caption{Approximated probability distribution of $\gamma$-ray energies for
radioactive decays of r-process material, guided by the early time $\gamma$-ray
emission spectrum by \citet{barnes2016a}. This includes 115 energy bins, which are sampled to obtain $\gamma$ energies
used in this work.
}

\label{fig:gammaenergies}
\end{figure}

We include $\gamma$-ray transport in our simulations,
as described by \citet{sim2007b}.
We approximate the emission energies of the $\gamma$-rays guided by the
early time $\gamma$-ray emission spectra obtained by
\citet{barnes2016a} (see their figure 3),
which peaks at several hundred keV.
The exact energies sampled and the probability of injecting a gamma-packet at each energy are shown in Figure~\ref{fig:gammaenergies}.

\section{Results}

\subsection{Dynamical ejecta}

\subsubsection{Bolometric light curves}
\label{sec:lightcurves}

\begin{figure}

 \includegraphics[width=0.5\textwidth]{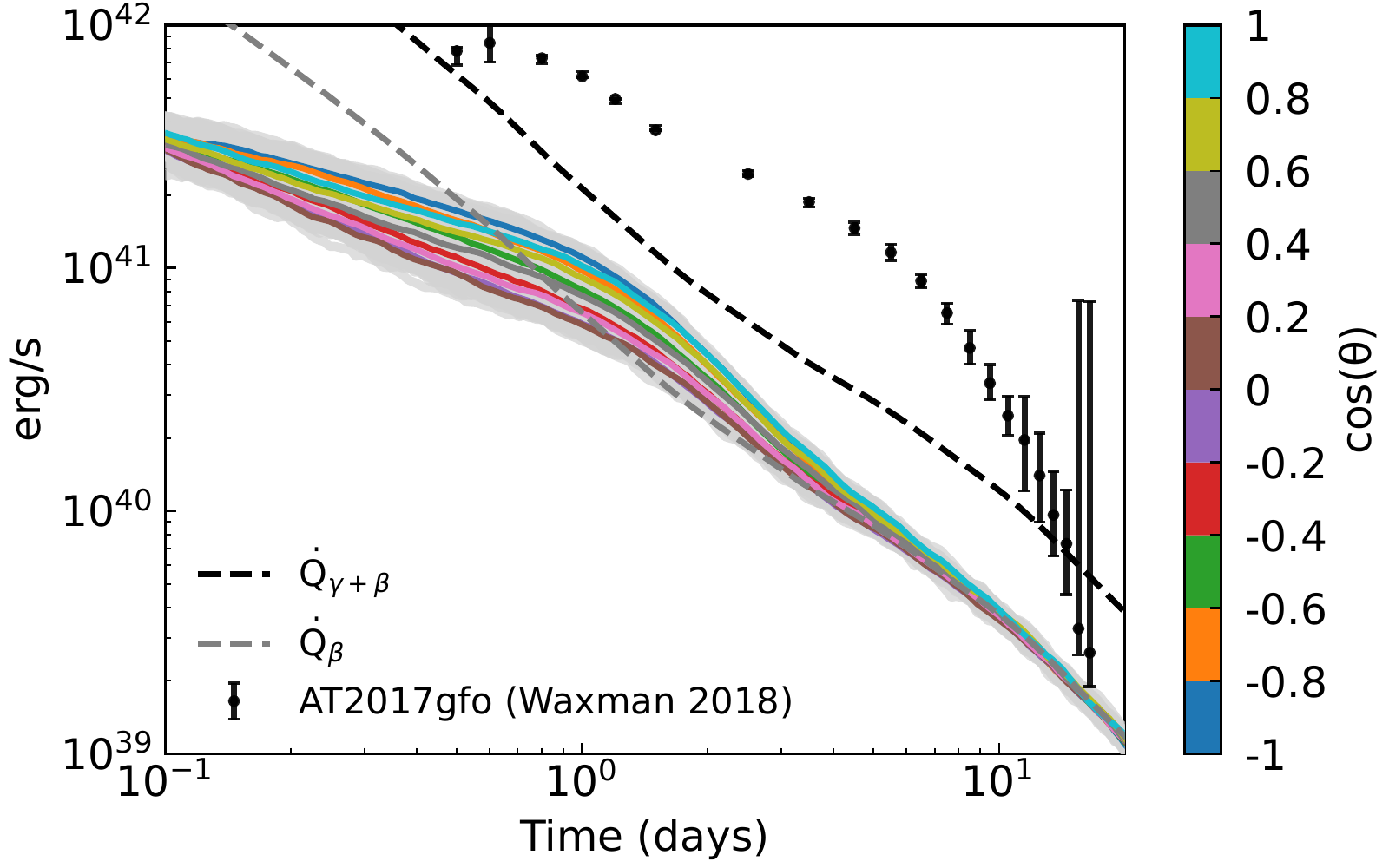}
 
 \includegraphics[width=0.5\textwidth]{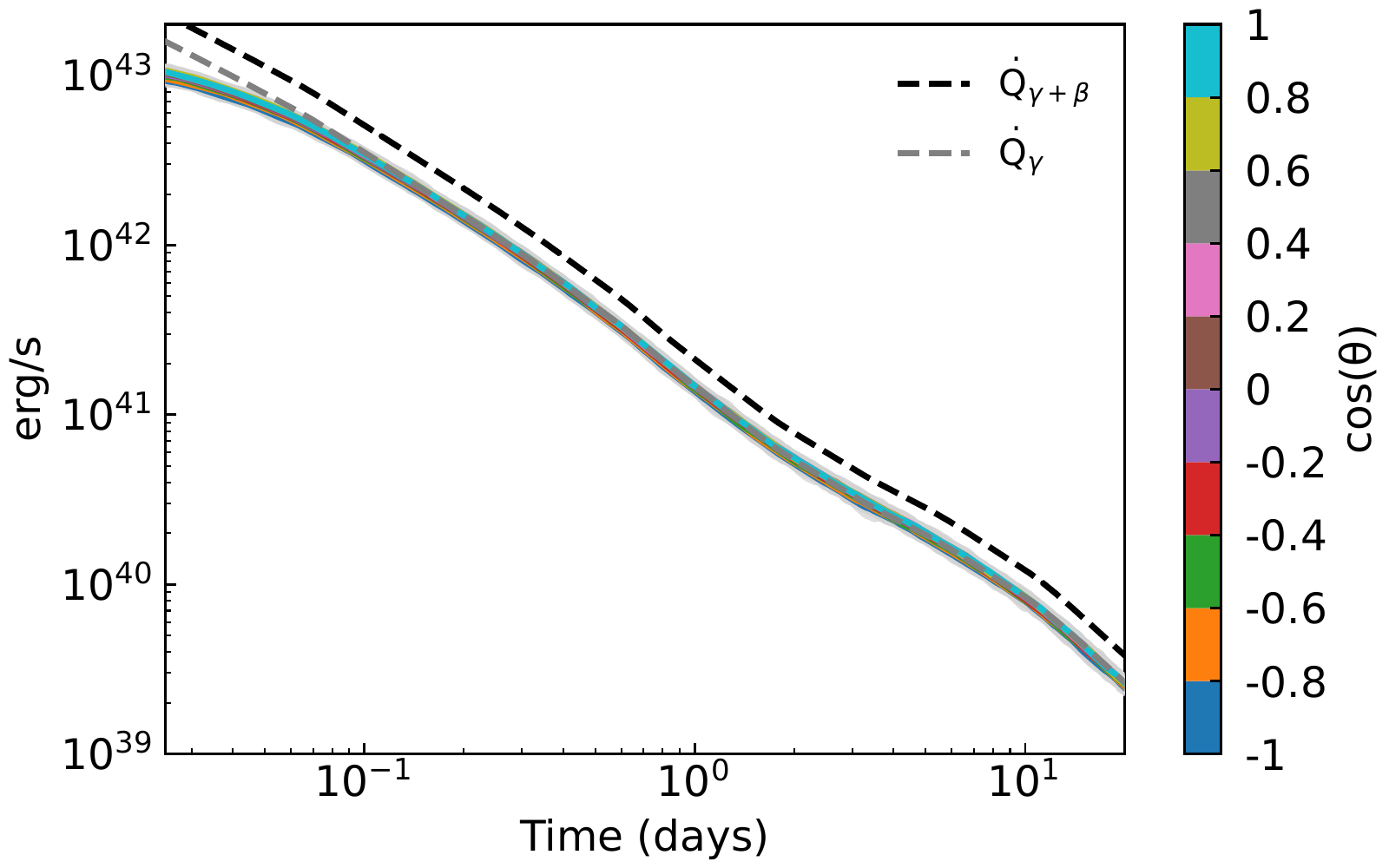}

\caption{Line of sight bolometric UVOIR light curves (upper panel)
and $\gamma$-ray light curves (lower panel).
Each coloured line is the azimuthally averaged light curve
(averaged over each angle bin in cos($\theta$)), while all 100 uniformly spaced
viewing angle bins are plotted in light grey.
When presenting the angle-dependent light curves, we plot them as equivalent isotropic luminosities (i.e. from the simulation we record the energy emitted per second into each {solid} angle bin to obtain light curves in  erg/s/sr for each orientation; we then scale these to an equivalent isotropic luminosity by multiplication by $4\pi \cdot$sr).
Also marked is the total energy released (on average) by nuclear reactions, $\dot{\mathrm{Q}}_{\gamma + \beta}$,
excluding the energy assumed to be lost to neutrino emission (35\%).
In the upper panel we additionally plot the energy due to $\beta$-particles ($\dot{\mathrm{Q}}_{\beta}$)
and in the lower panel the heating due to $\gamma$-rays ($\dot{\mathrm{Q}}_{\gamma}$).
The bolometric light curve of AT2017gfo from \citet{waxman2018a} is plotted
with the bolometric light curves for reference.
}

\label{fig:lightcurves}
\end{figure}

\begin{figure}

 \includegraphics[width=0.5\textwidth]{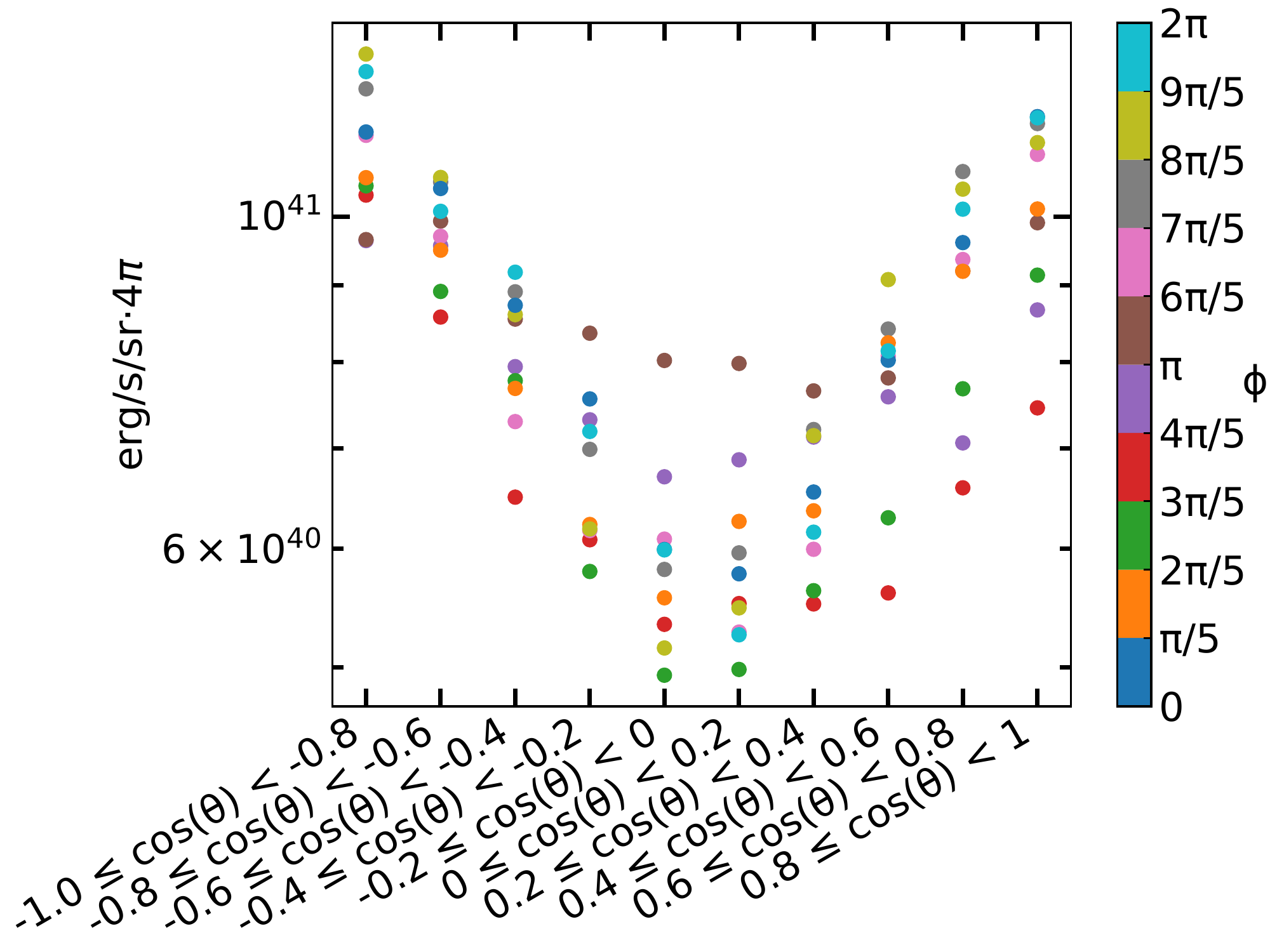}

\caption{Bolometric (UVOIR) equivalent isotropic luminosities in 100 uniform viewing angle bins at 1 day after the merger.
}

\label{fig:brightness1day}
\end{figure}

The bolometric light curves for the dynamical ejecta are shown in Figure~\ref{fig:lightcurves}.
From very early times photons are able to escape the ejecta,
hence we find that the light curves begin declining within a fraction of a day,
and do not show a significant rising phase in our simulations.
According to the nucleosynthesis calculations, significant numbers
of $\beta$-decays will occur in the outer ejecta layers,
hence, in our simulations, energy is deposited
in low density, low opacity outer regions, where it thermalises,
and is emitted as optical radiation from very early times.
Since we assume that all $\beta$-particle energy thermalises,
this may overestimate the true amount of energy that would thermalise,
however, at such times the thermalisation efficiency is likely to be high \citep{barnes2016a}.
The lack of rise to peak in bolometric light curves has previously been
found, e.g. by \citet{banerjee2020a}, \citet{klion2022a} and \citet{kawaguchi2022a}.
The light curves do, however, show a `shoulder' at $\sim 1$ day,
particularly in the polar directions.
At this time, the ejecta are becoming optically thin, and energy stored in the ejecta is able to escape. 
Photons are preferentially emitted in the polar
directions due to the lower optical depths
in these lines of sight (see Figure~\ref{fig:lightcurves}).
At times around 1 day after the merger, we find the strongest angle variation
in the light curves (see Figure~\ref{fig:brightness1day}), however this decreases over time
as the ejecta become optically thin.
As seen in Figure~\ref{fig:brightness1day}, the light curve also exhibits $\sim 30\%$ variations in azimuthal viewing angle,
likely reflecting that the ejecta do not show perfect cylindrical symmetry.

The heating rate marked in Figure~\ref{fig:lightcurves} shows the 
angle-averaged, total
amount of energy available for heating the ejecta over time.
Since we assume all energy comes from $\beta$-decays in
our simulations, the heating rate represents
the total energy from
$\gamma$-rays plus $\beta$-particles from $\beta$-decays.
The energy lost to neutrino emission is excluded
from the heating rate shown in Figure~\ref{fig:lightcurves}.
Since $\gamma$-rays thermalise inefficiently (see Section~\ref{sec:gammalightcurves})
it is predominantly energy from $\beta$-particles
powering the light curve.
By $\sim 5$ days the ejecta have become optically thin,
and the light curve is equal to the $\beta$-particle heating rate.
At this time $\beta$-particles are no longer expected
to thermalise efficiently \citep[e.g.][although we note this was for a simplified ejecta structure]{barnes2016a}
and so the $\beta$-particle heating rate is likely overestimated
here. 
However, energy from $\alpha$-particles becomes more
significant at these times.
Since the late time light curve is dependent on the assumed heating
rate, this highlights the importance of calculating
the fractions of $\beta$-decay energy going into 
neutrinos, $\gamma$-rays and $\beta$-particles,
by following the radioactive decays 
of r-process elements in the radiative
transfer simulations,
as well as calculating how much of that energy will thermalise.
Accounting for $\alpha$-particles, and the corresponding thermalisation efficiency of the decay products
is likely also important at late times.

For reference, the bolometric light curve of AT2017gfo,
constructed by \citet{waxman2018a}, is also plotted in Figure~\ref{fig:lightcurves}.
The mass of the dynamical ejecta in our model
is relatively small ($\sim$10 times less than e.g. the mass inferred for AT2017gfo by \citealt{smartt2017a}, using an Arnett-type model, of $0.04 \pm 0.01$ \msun)
and therefore the total amount of energy produced in
our model is lower, leading to fainter, faster evolving light curves.

{We note that due to r-process heating, ejecta velocities may be increased by the additional energy released. We test the effect of this on the light curves in Appendix~\ref{sec:velocityboost}.}

\subsubsection{$\gamma$-ray light curves}
\label{sec:gammalightcurves}

We also show the emerging $\gamma$-ray light curves in
each viewing angle bin in Figure~\ref{fig:lightcurves}.
For all except the earliest times, the $\gamma$-rays
do not thermalise and are able to free-stream out of the
ejecta.
By $\sim 0.1$ days the emerging $\gamma$-ray energy is equal
to the total $\gamma$-ray energy rate, marked in Figure~\ref{fig:lightcurves}.
The $\gamma$-ray light curves do not show a
viewing angle dependence.
To observe the peak of the $\gamma$-ray light curve, observations would
need to be within the first hour after the merger.

\begin{figure}
\centering

\includegraphics[width=0.5\textwidth]{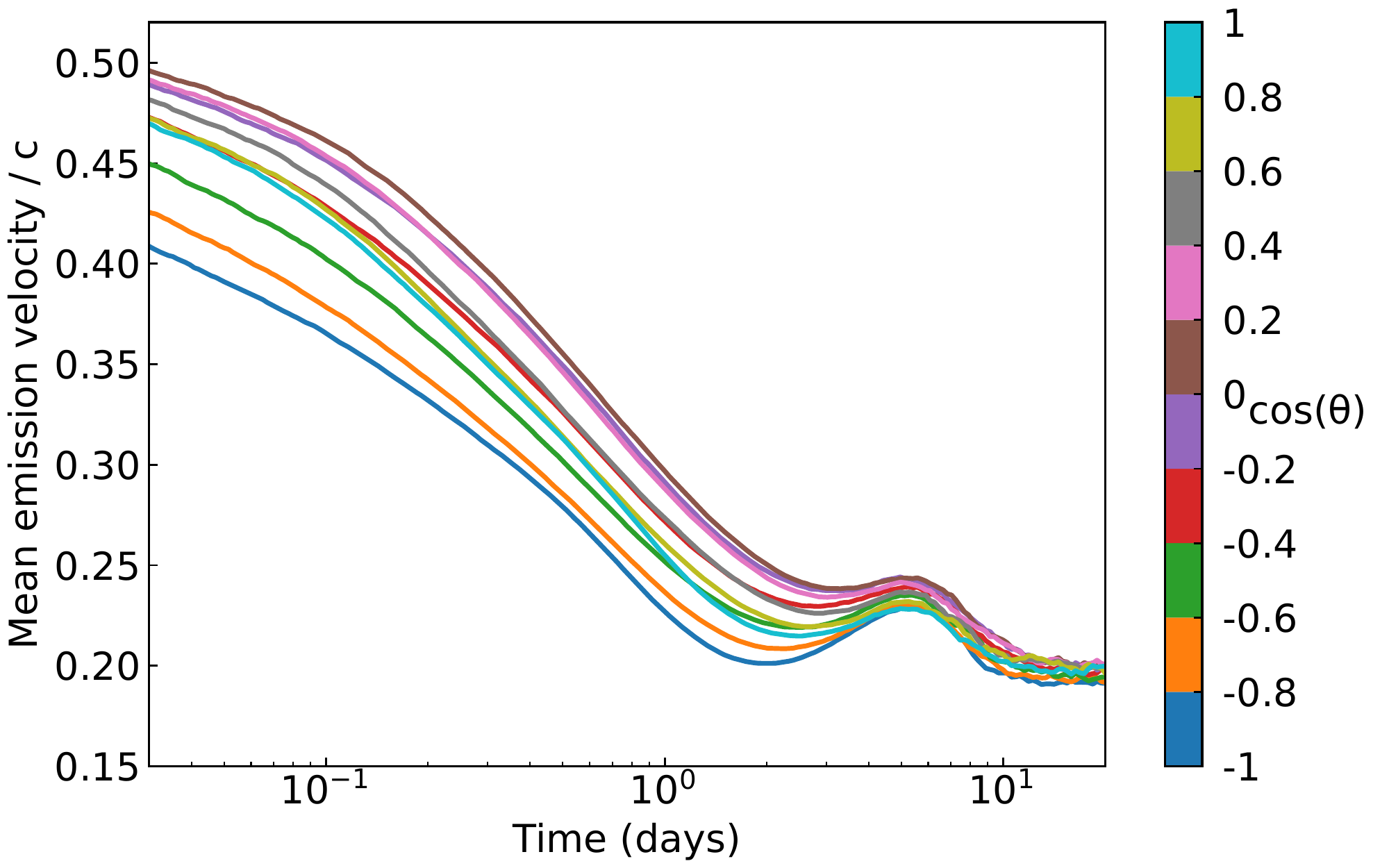}

\caption{Mean ejecta velocity at which Monte Carlo packets
were last emitted within a given angle bin, indicating the ejecta velocity where packets last interacted.
}

\label{fig:emission_velocity_angle_ave}
\end{figure}

\begin{figure*}
     \begin{subfigure}[b]{0.35\textwidth}
         \centering
            \includegraphics[width=\textwidth]{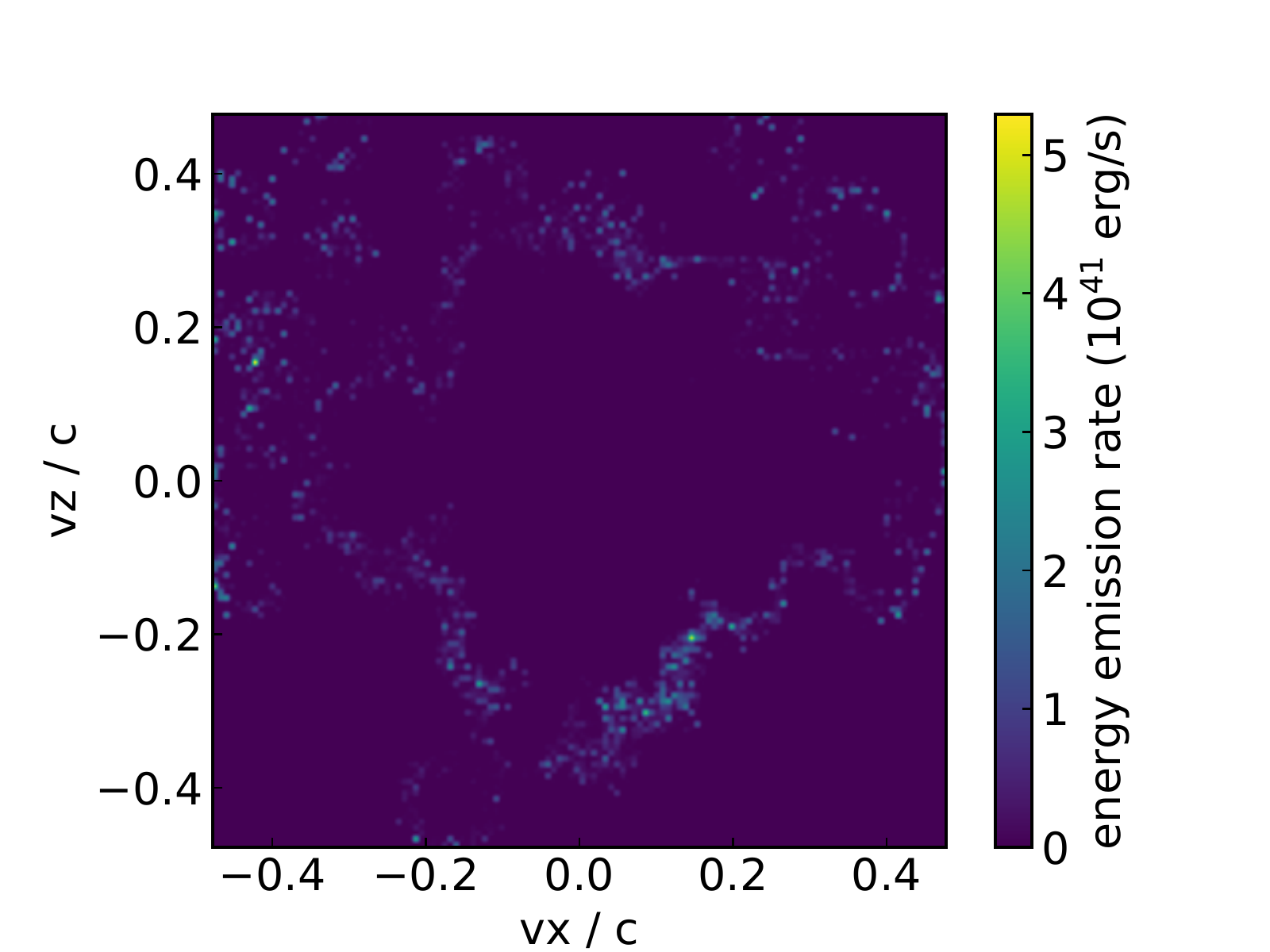}
            \caption{0.2 days}
    \end{subfigure}
     \begin{subfigure}[b]{0.35\textwidth}
         \centering
            \includegraphics[width=\textwidth]{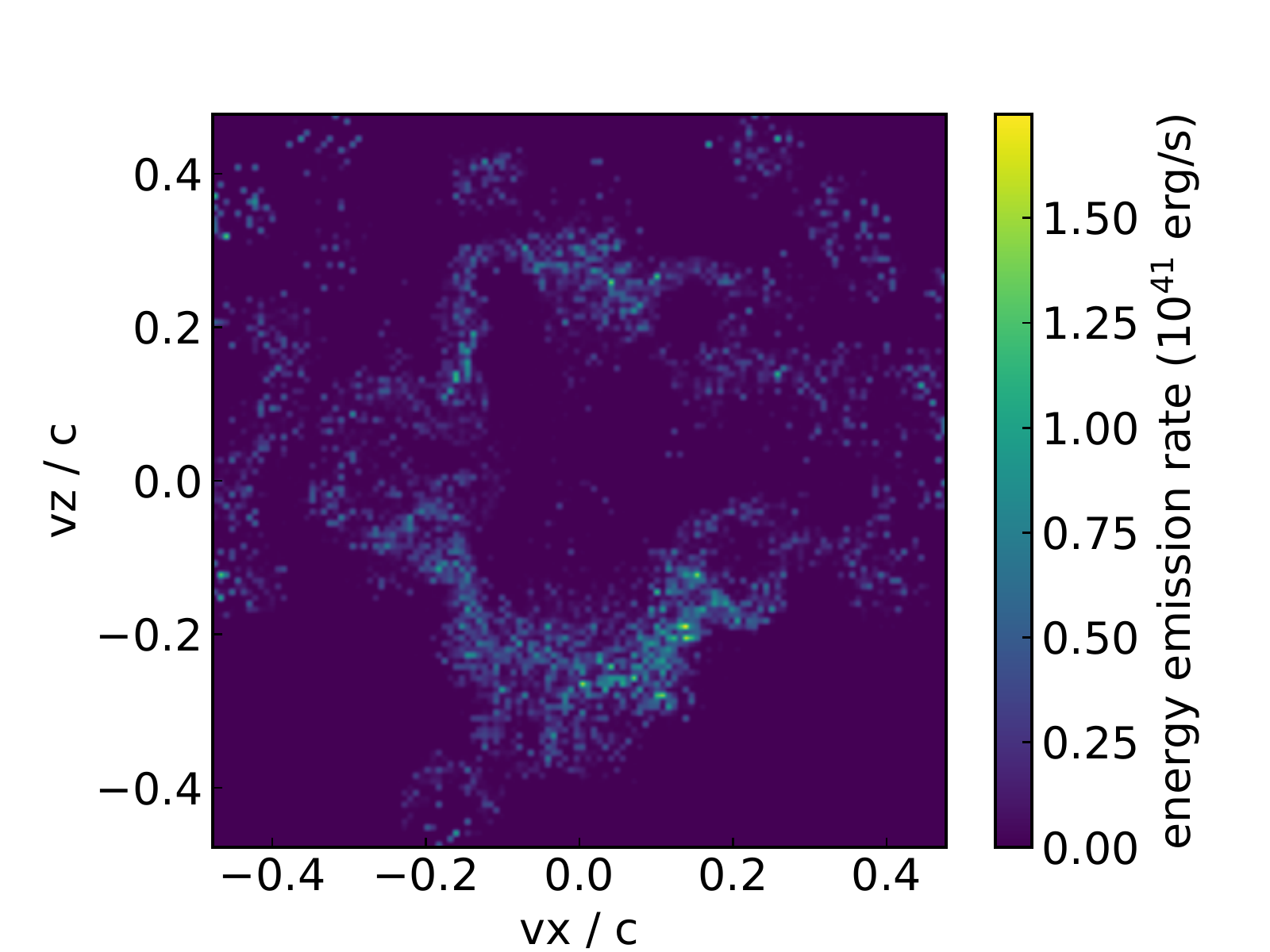}
            \caption{0.5 days}
    \end{subfigure}

     \begin{subfigure}[b]{0.35\textwidth}
         \centering
            \includegraphics[width=\textwidth]{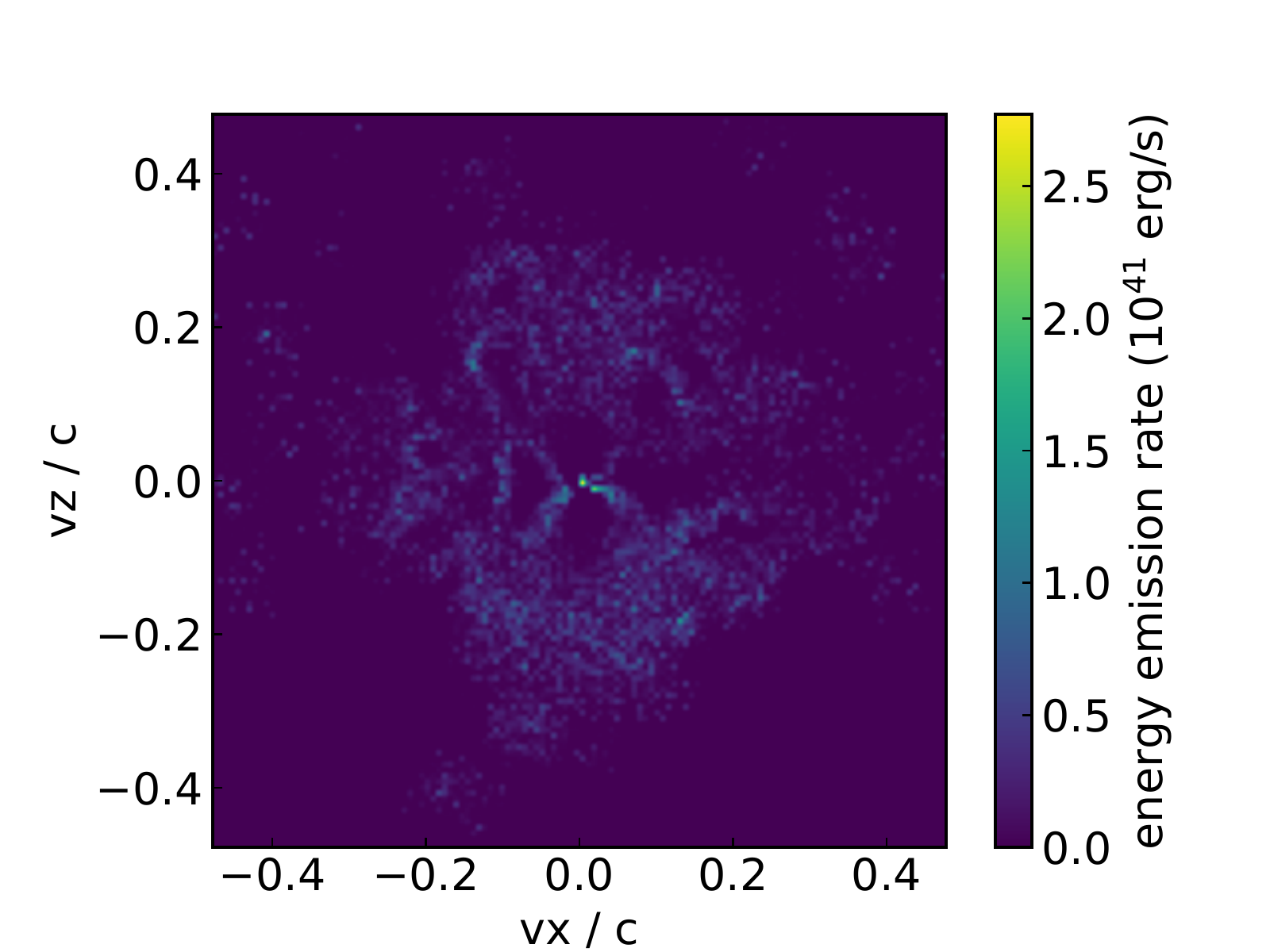}
            \caption{1 day}
    \end{subfigure}
     \begin{subfigure}[b]{0.35\textwidth}
         \centering
            \includegraphics[width=\textwidth]{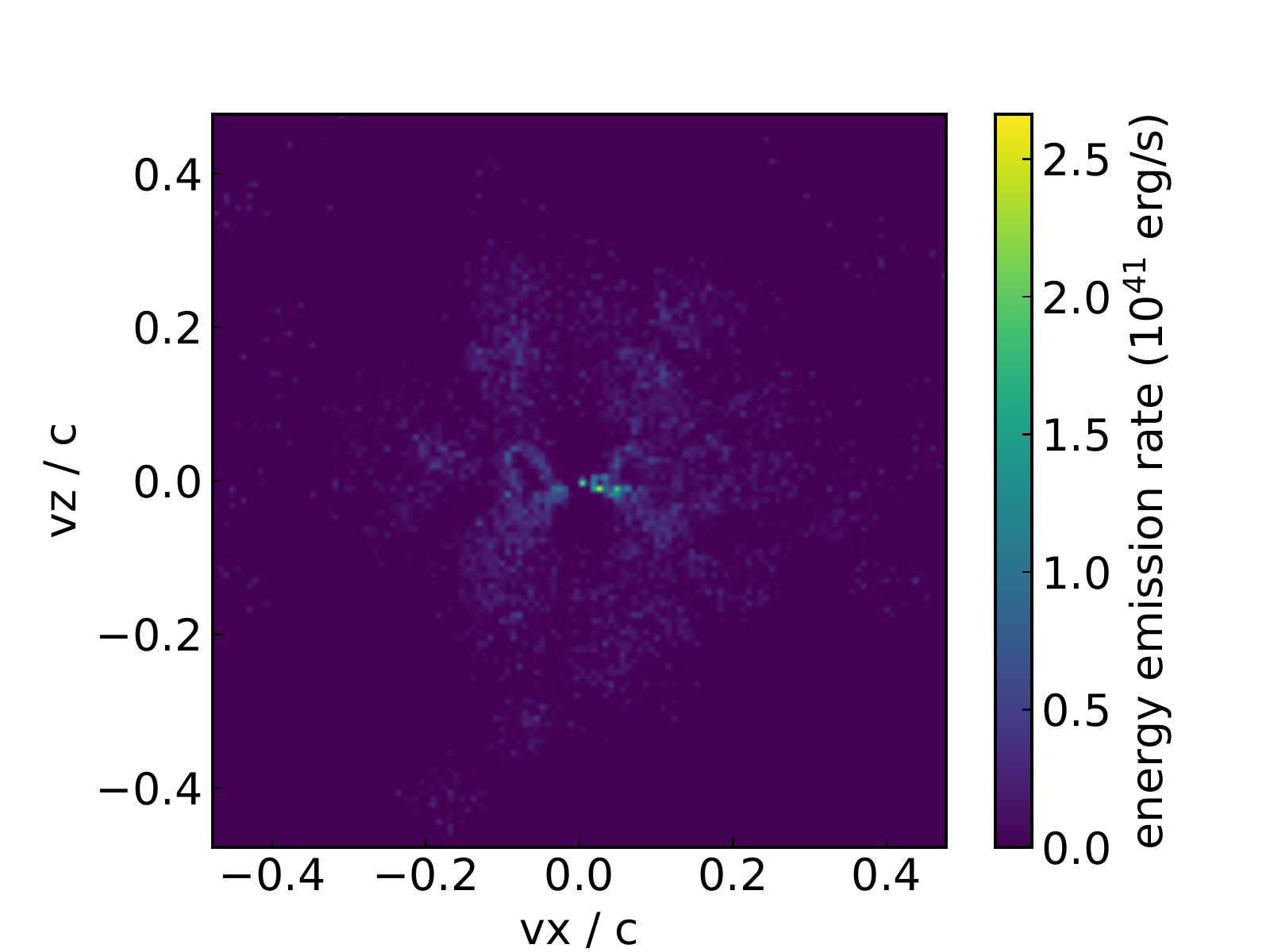}
            \caption{2 days}
    \end{subfigure}

\caption{The spatial distributions (slices at v$_{\rm y} = 0$) of last interactions of escaping radiation at three epochs.
Note packets can escape in any direction -- we make no selection based on escaping angle.
Packets that escaped representing $\gamma$-rays
are excluded from this.
}

\label{fig:emission_velocity_grid}
\end{figure*}

\subsubsection{Ejecta emission velocities}

As discussed in Section~\ref{sec:lightcurves},
initially most of the photon emission occurs
in the outer ejecta layers,
hence we find that the bolometric light curves begin declining from
very early times and do not rise to a peak within the time frame of our simulation.
Figure~\ref{fig:emission_velocity_angle_ave}
shows the mean ejecta velocity at which
escaping Monte Carlo packets underwent their last interaction in the simulation -- this gives an indication of the regions of the ejecta that are contributing to the kilonova emission.
The outer ejecta layers quickly become optically thin,
due to the high expansion velocities.
As a result, the ejecta velocities from which packets are emitted decrease rapidly 
within the first day.
We show the mean ejecta velocities from which packets are emitted
in Figure~\ref{fig:emission_velocity_angle_ave}.
{Since radioactive material is synthesised throughout the ejecta, and the energy distribution does not peak steeply, at late times when the ejecta are optically thin the emission velocity indicated in Figure~\ref{fig:emission_velocity_angle_ave} is an average of the radial velocities at which radioactive decays occur and energy thermalises.
The mass-weighted mean velocity of the ejecta is 0.2c, which corresponds to the mean emission velocity once the ejecta are optically thin.
We would expect the energy distribution (and density distribution) to peak more steeply at lower velocities when a secular ejecta component is included. In this case the mean emission velocities when the ejecta are optically thin are expected to be lower (see Section~\ref{sec:mean_emission_velocity_secular}).}

We also show slices of the ejecta indicating where the
Monte Carlo packets escape from at a given time in Figure~\ref{fig:emission_velocity_grid}.
Initially, all packets are emitted from the outermost (i.e. fastest) ejecta.
Until around 1 day, this resembles a photosphere, beneath which radiation
does not escape the ejecta.

\subsubsection{Temperature}
\label{sec:temperatures}

\begin{figure}
\centering

\includegraphics[width=0.5\textwidth]{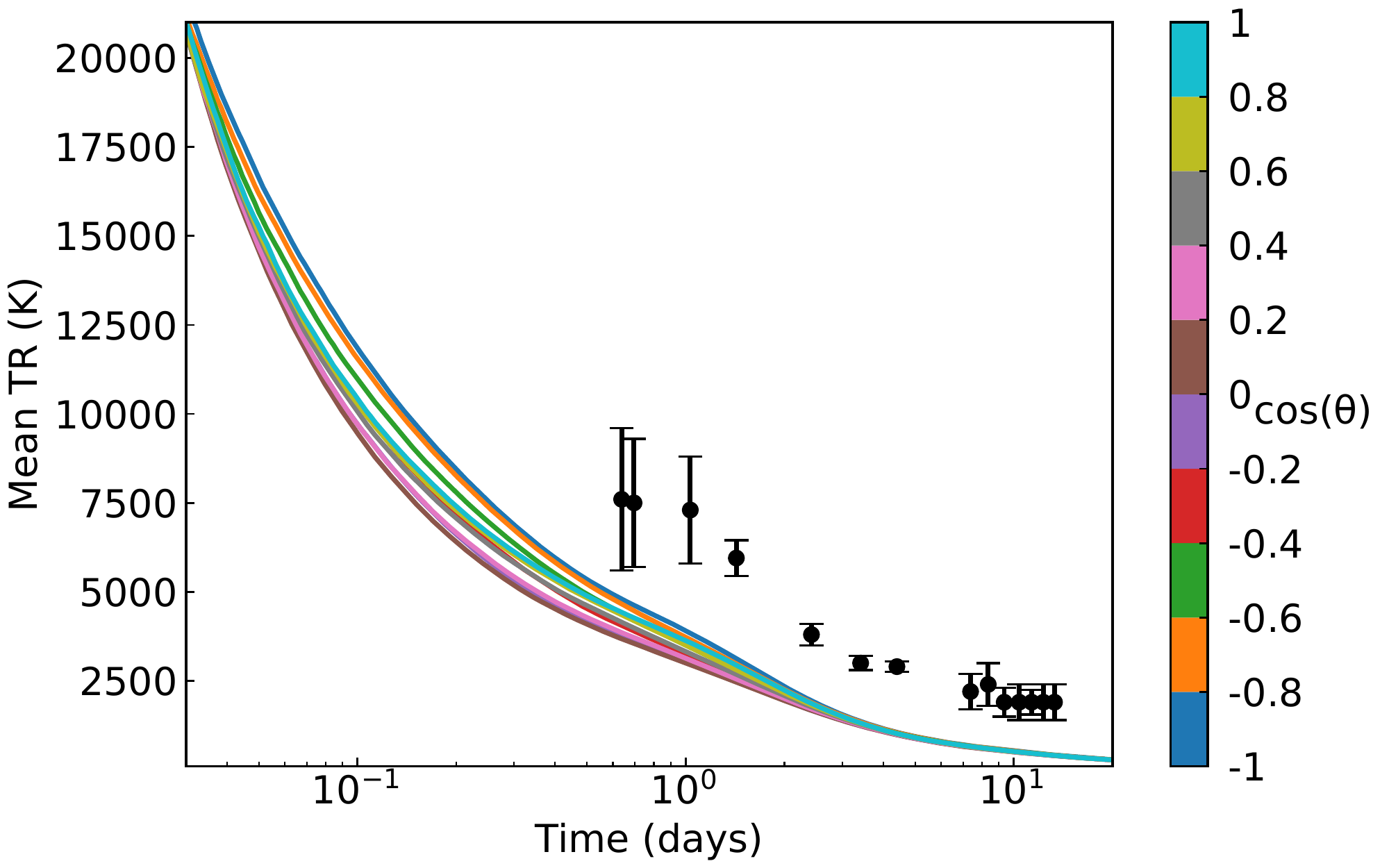}

\caption{
Average radiation temperature at the last interaction locations of escaping packets, indicating the ejecta temperatures where radiation is escaping from.
We also mark the inferred temperatures from the spectra of AT2017gfo by \citet{smartt2017a}.
}

\label{fig:emission_velocity_temperatures}
\end{figure}

\begin{figure*}

\includegraphics[width=0.9\textwidth]{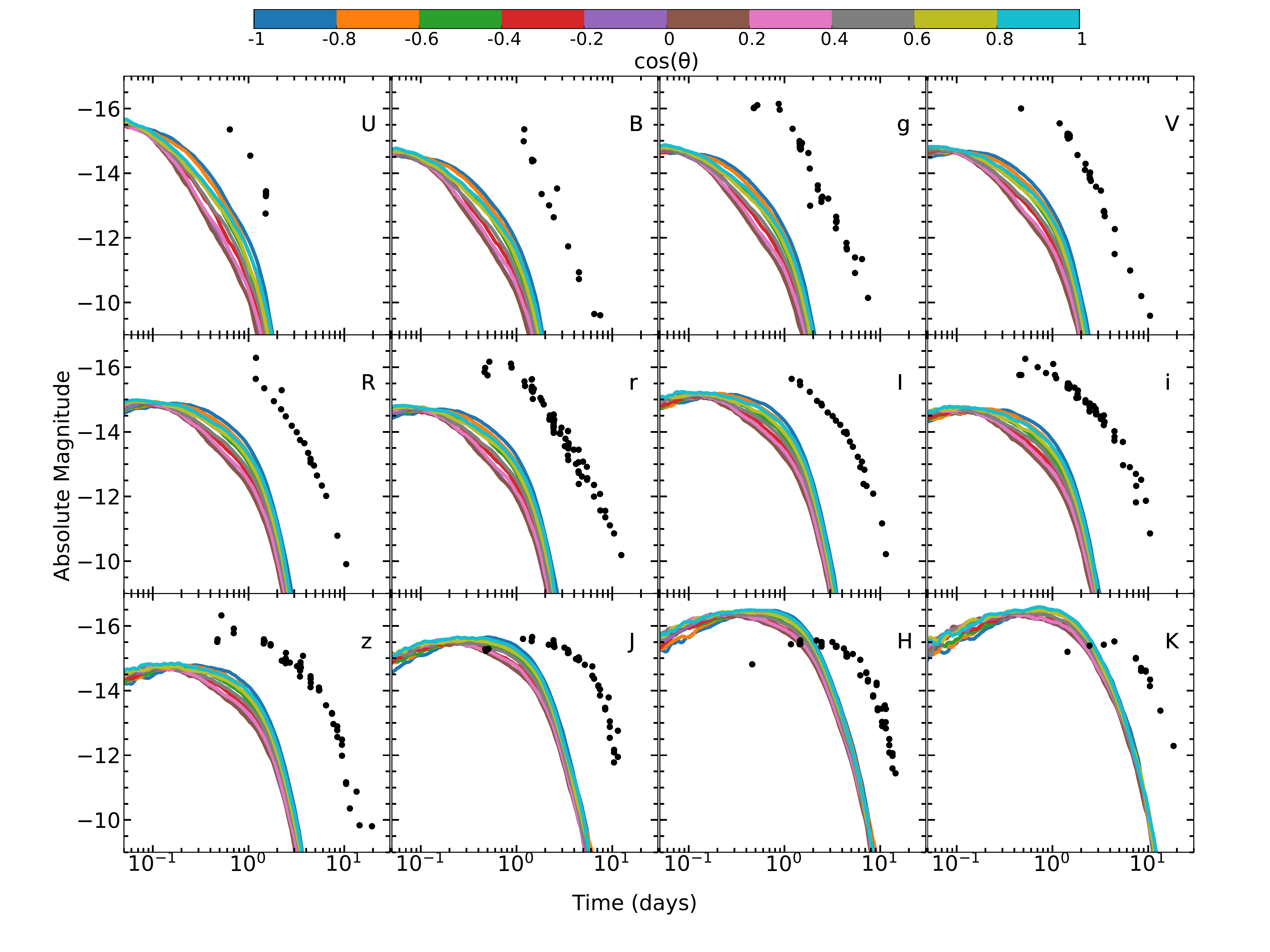}

\caption{Light curves estimated from radiation temperature.
Also plotted are the light curves for AT2017gfo from \citet{villar2017a}, assuming a distance of 40 Mpc.
We note that the light curves show some Monte Carlo noise. 
}

\label{fig:bandlightcurves}
\end{figure*}

The extremely high expansion velocities cause the temperature of the ejecta to drop rapidly.
We estimate the temperature in each model grid cell
by matching the energy density of the Monte Carlo radiation field
in the cell during each time step to that of a black body radiation field:
we denote this equivalent black body temperature as T$\rm_J$.
We show the average temperatures of the model grid cells from
which packets are escaping
over time in Figure~\ref{fig:emission_velocity_temperatures}.
Specifically, we record the cell temperature where a packet last
interacted before escaping and bin the packets in time.
The mean temperature in each time bin is plotted.

The opacities we adopt from \citealt{tanaka2020a} (see Table~\ref{tab:opacities}) were calculated
for temperatures of 5000 - 10000 K.
In our simulations the temperatures are approximately in this range, at least in the regions from which packets escape, 
between 0.1 -- 1 days.

At late times it is likely that we underestimate the temperature,
due to the grey and local thermodynamic equilibrium (LTE) approximations assumed here.
In the nebular phase, non-thermal processes
would provide heating to the ejecta \citep{hotokezaka2021a, pognan2022a} and likely prevent cooling to such low temperatures as suggested by our T$_{\rm J}$ estimate.

We also plot the mean ejecta temperatures at the location
of the last interaction of escaping radiation
(specifically, the temperature of the model grid cell where a packet last interacted before escaping),
binned by time
in Figure~\ref{fig:emission_velocity_temperatures},
and we show how the temperatures compare to the inferred
temperatures of AT2017gfo by \citet{smartt2017a}.
Although the model temperatures are cooler than
AT2017gfo by the time of the observations,
the evolution of the temperature is similar.
The cooler temperatures are likely due to the lower
mass of the ejecta model, and therefore the lower amount of energy generated in the ejecta,
as well the LTE assumptions made in our simulation.

\subsubsection{Colour curves from radiation temperature}

\begin{figure}

\includegraphics[width=0.45\textwidth]{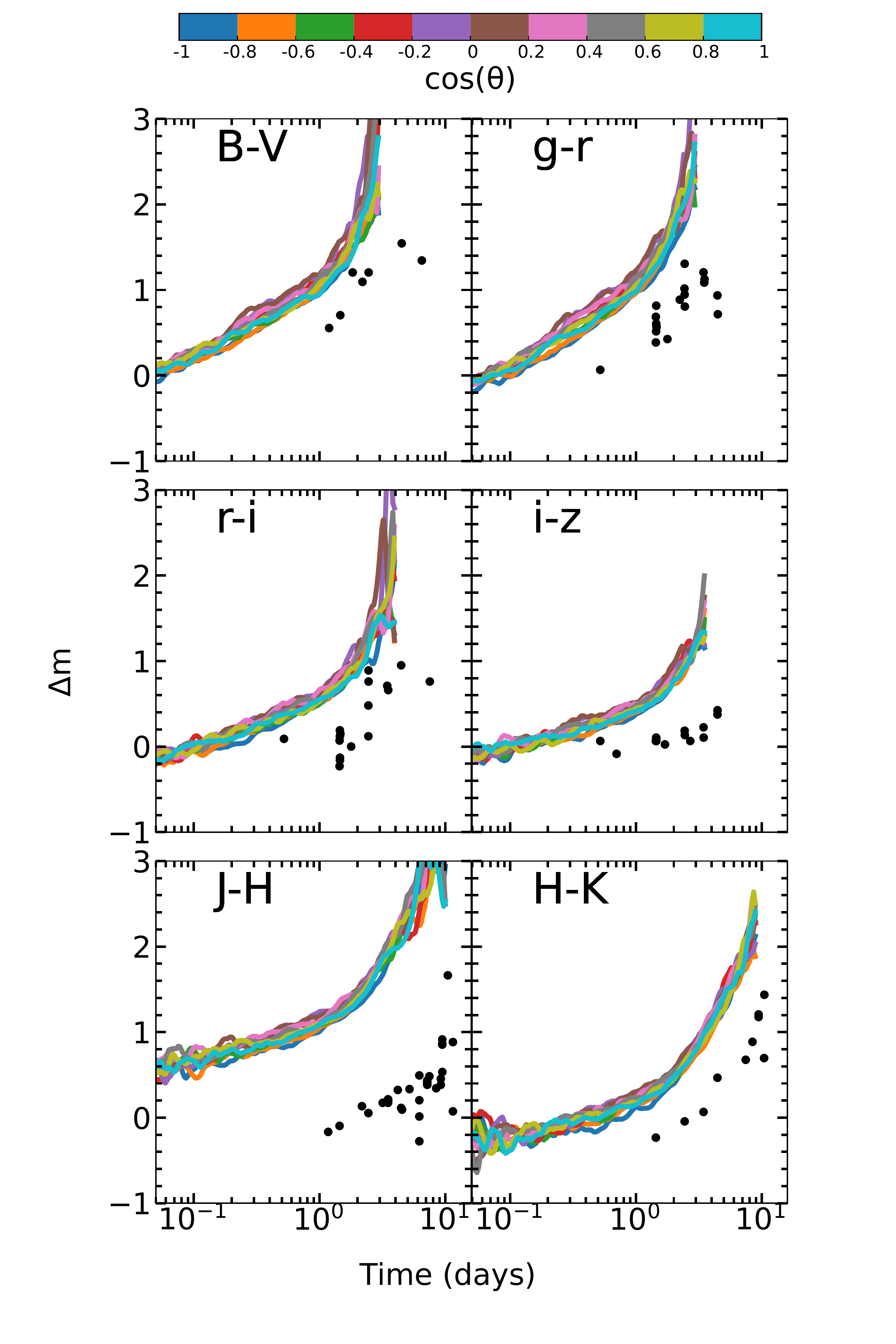}

\caption{Angle-averaged colour evolution, estimated from radiation temperature. Also plotted are the
colours of AT2017gfo for points where observations were taken at the same time
    (i.e. we do not interpolate between light curve points).
    Note that Monte Carlo noise can be seen here.}

\label{fig:colourevolution}
\end{figure}

Our simulations use only grey opacities, so we do not
calculate any frequency dependence to produce
spectra or colour information from the Monte Carlo packets directly.
However, we record the temperature
of the grid cell where each packet
was last emitted from before escaping the
ejecta, i.e. the point at which the packet last
underwent an interaction (see Figure~\ref{fig:emission_velocity_temperatures}).
From this, we assign an approximate frequency to
the Monte Carlo packet by making the simplistic assumption that it is governed by the equilibrium radiation distribution (a black body function)
at the local radiation temperature, T$\rm_R$ (T$\rm_R$ = T$\rm_J$ under the LTE assumption).
We do this by randomly sampling a frequency from a
black body at the temperature in the model grid cell from which the packet escaped.
From this we obtain an approximate spectral energy distribution,
and can generate band-limited light curves,
which we show in Figure~\ref{fig:bandlightcurves}.

The elemental compositions of the kilonova ejecta are dominated by lanthanides and actinides,
which are very effective at line-blocking blue wavelengths,
and therefore we would expect that the true spectra would be redder
than a pure black body, particularly at later times \citep[e.g.][]{gillanders2022a}.
However, we would also expect the temperature evolution
to change in non-grey and non-LTE simulations, which would also affect
the colour evolution.
To accurately determine band-limited light curves,
frequency dependent simulations are required, which goes beyond the scope
of this work.

The light curves show an angle variation in magnitude of up to 
$\sim$1~mag.
The emission from polar directions is brighter than from equatorial directions.
As discussed in Section~\ref{sec:model},
we have a higher $Y_{e}$, and therefore lower lanthanide fraction
in the polar directions,
which represents a `blue' component,
and a lower $Y_{e}$ in the direction of the equator,
which would lead to a higher lanthanide fraction,
representing a `red' component.
However, this does not lead to significantly
redder colours in the equatorial direction than the polar directions.
We show the angle dependent colour evolution in Figure~\ref{fig:colourevolution}, which does not show a significant
angle variation. 

Due to the evolution of the temperature alone, we find that
the colours show a rapid evolution from blue to red over time
in all lines of sight,
which can be seen in Figure~\ref{fig:colourevolution}.
{At early times when the radiation temperatures are high (see Figure~\ref{fig:emission_velocity_temperatures}) relatively blue frequencies are sampled. At later times when the ejecta have cooled the black body peaks towards the red, leading to redder frequencies being sampled.
The rapid cooling of the ejecta due to the high expansion velocities drives the rapid colour evolution found in Figure~\ref{fig:colourevolution}.}

\subsubsection{Comparison to AT2017gfo}

We plot the light curves of AT2017gfo \citep[from][]{villar2017a} in Figure~\ref{fig:bandlightcurves}
for reference,
although again we note that the model we consider here is much
less massive than what was inferred for AT2017gfo, and therefore
we do not expect to match the brightness.
The light curves have been corrected for foreground reddening,
assuming an extinction of E(B-V) = 0.11 mag \citep{smartt2017a}.

Interestingly, the model light curves show a similar evolution
to AT2017gfo, although the model light curves are fainter
and evolve faster.
This could suggest that in future events where less mass is
ejected than in the case of AT2017gfo, the light curves
would show a faster decline and may have already faded
in the bluer bands by the time the first detections
of AT2017gfo were made.
The red bands, however, remain brighter for longer, suggesting that
searches for the electromagnetic counterparts of future kilonova events should
focus on these bands.
We note that the model H and K band light curves are likely too red,
since these are similar in brightness to AT2017gfo.
As discussed in Section~\ref{sec:temperatures},
due to our grey approximation we likely underestimate the
temperature at later times, which could be responsible for the
very red late-time colours.

The rapid blue to red colour evolution found for our model is similar to that observed for AT2017gfo (see Figure~\ref{fig:colourevolution}).
Since we only include dynamical ejecta in this simulation,
this shows that a late-time, high opacity component is not necessarily
required to explain the colour evolution shown by AT2017gfo,
as has been found by e.g. \citet{cowperthwaite2017a},
although we note that we do not claim there was no secondary component.
This suggests that the colour evolution could be driven by the cooling of the ejecta.
Future frequency dependent simulations will be required to
confirm this.
{We note that previous studies have also found that single component models can explain the observations of AT2017gfo, including \citet{tanaka2017a} and \citet{waxman2018a}.}

\subsection{Late-time ejecta}

So far, we have only considered the dynamical ejecta,
expelled on timescales of tens of milliseconds.
Now we investigate in an approximate manner the effects of ejecta expelled on longer timescales.

\subsubsection{Model for late-time ejecta}

\begin{figure}

\includegraphics[width=0.5\textwidth]{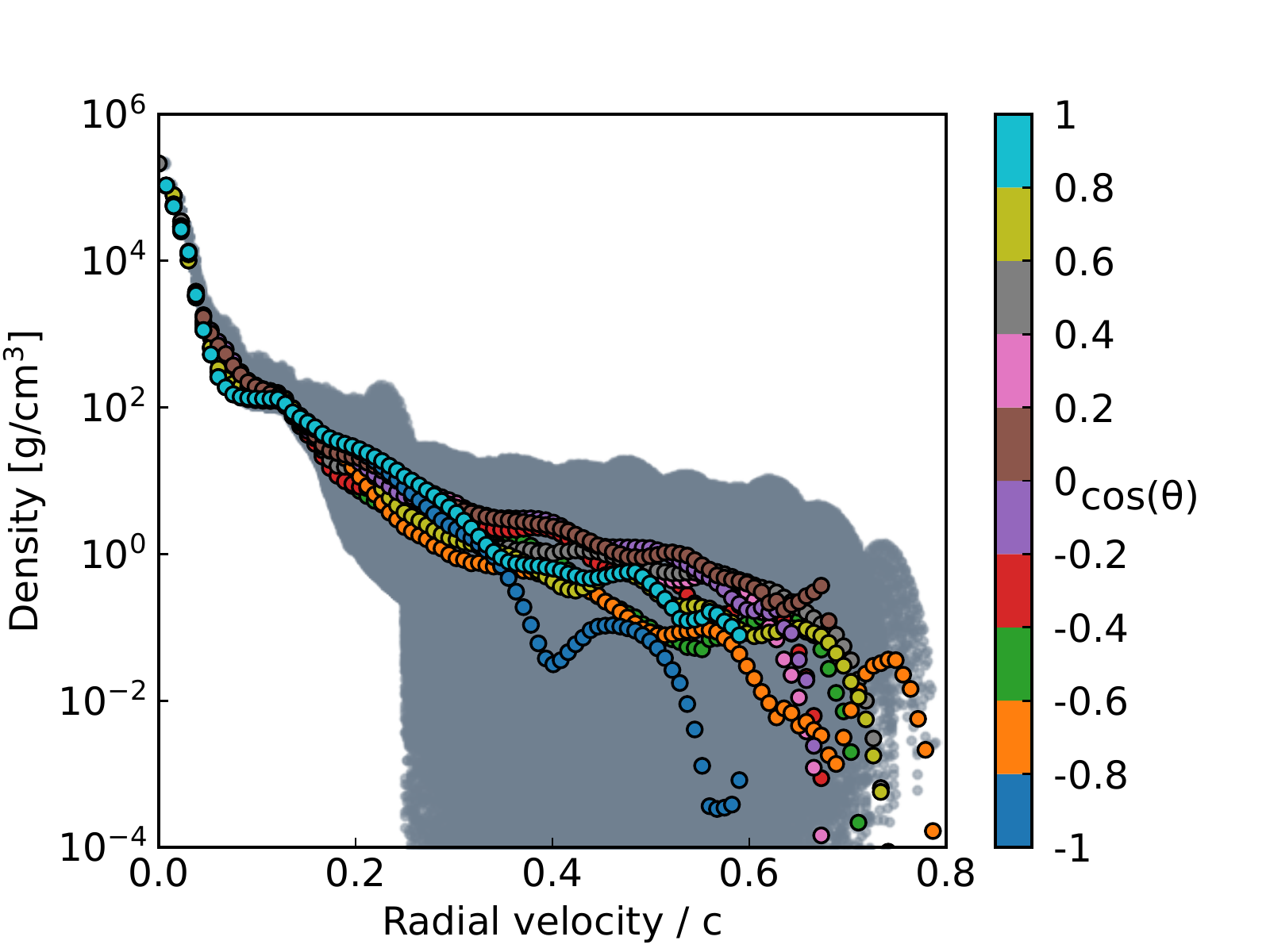}

\caption{Same as Figure~\ref{fig:rho-everycell} but including the spherically averaged late-time ejecta. The densities of the late-time ejecta have been scaled (assuming homologous expansion) to match the time in Figure~\ref{fig:rho-everycell} of 0.52 seconds.
}

\label{fig:cell_densities_plus_late_ejecta}
\end{figure}

We base the density structure for the secular ejecta on
a simulation of a BH-accretion torus, which is formed after the collapse of the hypermassive neutron star.
The BH-torus model is similar to the models
of \citet{just2015a}, but includes a special relativistic treatment as well as additional, manually constructed ejecta components.
The components represent dynamical ejecta and a neutrino-driven wind from a hyper-massive neutron star (HMNS) as described by \citet[][see therein for more details]{ito2021a}. 
The initial disk mass was 0.1 M$_\sun$, the mass of the HMNS wind was 0.006 M$_\sun$, the black hole mass was 2.7 M$_\sun$, and the assumed time of BH formation after the merger was 0.05 s.
The torus and wind components from the long term evolution simulation
are spherically averaged and then mapped to the 3D Cartesian grid (i.e. the dynamical ejecta component from this simulation is neglected).
In each cell of our Cartesian grid, we add the corresponding mass from the remnant simulation to the mass of the dynamical ejecta,
described in Section~\ref{sec:model}.
We do not use the $Y_{e}$ information from the remnant model, but keep the $Y_{e}$ and energy (per unit of mass) distributions
that were determined for the dynamical ejecta.
Any empty cells in the original dynamical ejecta model without a $Y_{e}$
(close to the centre where the bound remnant was removed)
are given a $Y_{e}$ of 0.5
and a total heating energy equal to the average heating rate of the 
dynamical ejecta trajectories
(marked in Figure~\ref{fig:trajectoriesQdot}) integrated over time.
The additional mass of the late-time ejecta is 0.019 \msun,
which gives a total ejecta mass of 0.024 \msun.

In Figure~\ref{fig:cell_densities_plus_late_ejecta} we show
the model cell densities after including the mass
from the torus and wind ejecta.
Due to the very high central densities introduced by the
late-time ejecta model, significantly higher amounts of energy will be generated in the low velocity, central regions of the ejecta (since energy released is proportional to mass).
We therefore expect the light curve to be brighter
at late times when this energy diffuses out of the ejecta.

\subsubsection{Effect on light curves of late-time ejecta}
\label{sec:late-time-ejecta-lightcurves}

\begin{figure}
\includegraphics[width=0.45\textwidth]{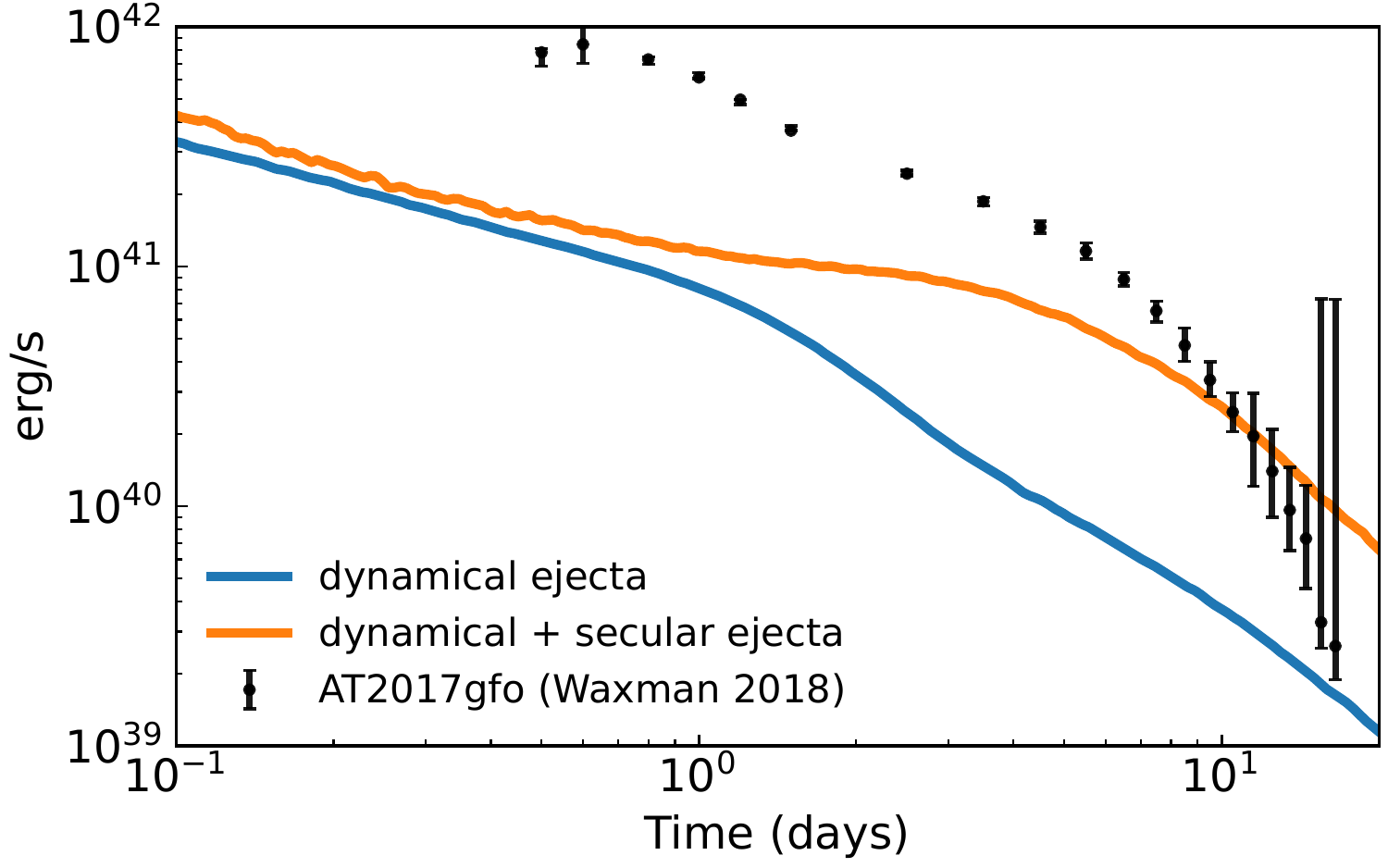}

\caption{Bolometric angle-averaged light curves for the model including the secular ejecta component and without.
We note that Monte Carlo noise can be seen in this plot.
}

\label{fig:lightcurves-angleave-secularejecta}
\end{figure}

We carry out a radiative transfer simulation on the model combining the late-time ejecta with the dynamical ejecta. 
This simulation was carried out between 0.05 and 120 days with $2.88 \times 10^7$ Monte Carlo packets.

To demonstrate the effect of the late-time ejecta on the 
bolometric light curve in comparison to the dynamical ejecta alone,
we plot the angle-averaged bolometric light curves of
each model
in Figure~\ref{fig:lightcurves-angleave-secularejecta}.
Due to the greater mass in the center of this model,
the increased energy generated in the high density inner regions
leads to a peak (or shoulder) in the bolometric light curve at $\sim 5$ days
after the merger, which was not produced by the dynamical ejecta
alone.
The increased energy thermalising
at low velocities and high optical depths
takes longer to diffuse out of the ejecta,
and we find a later peak in the bolometric light curve.
This highlights the necessity of long term hydrodynamic simulations,
since the late-time light curve is dominated by the lower velocity,
late-time ejecta.
At the same time, the small difference between the two cases at early times ($\lesssim 1$ day) suggests that the early light curve
may be dominated by the dynamical ejecta.

At late times (>10 days) the model light curve shows a slower decline than the observations of AT2017gfo.
This is likely due to our assumption that all $\beta$-particle
energy will thermalise at all times, and suggests that by these times the actual thermalisation rate is lower than we assume.

The late time light curve is now a similar brightness to AT2017gfo,
suggesting the mass of our model at low velocities may be similar
to that of AT2017gfo.
However, the relative faintness at early times indicates that our model
would require more mass at higher velocities to match the brightness
of AT2017gfo.

{Since for this model we kept the Ye structure of the dynamical ejecta simulation, the opacity of the late-time ejecta likely does not match simulations.
For this reason we do not discuss the effect on the approximate band-limited light curves and colour evolution.}

\subsubsection{Effect on mean emission velocities}
\label{sec:mean_emission_velocity_secular}

\begin{figure}
\includegraphics[width=0.45\textwidth]{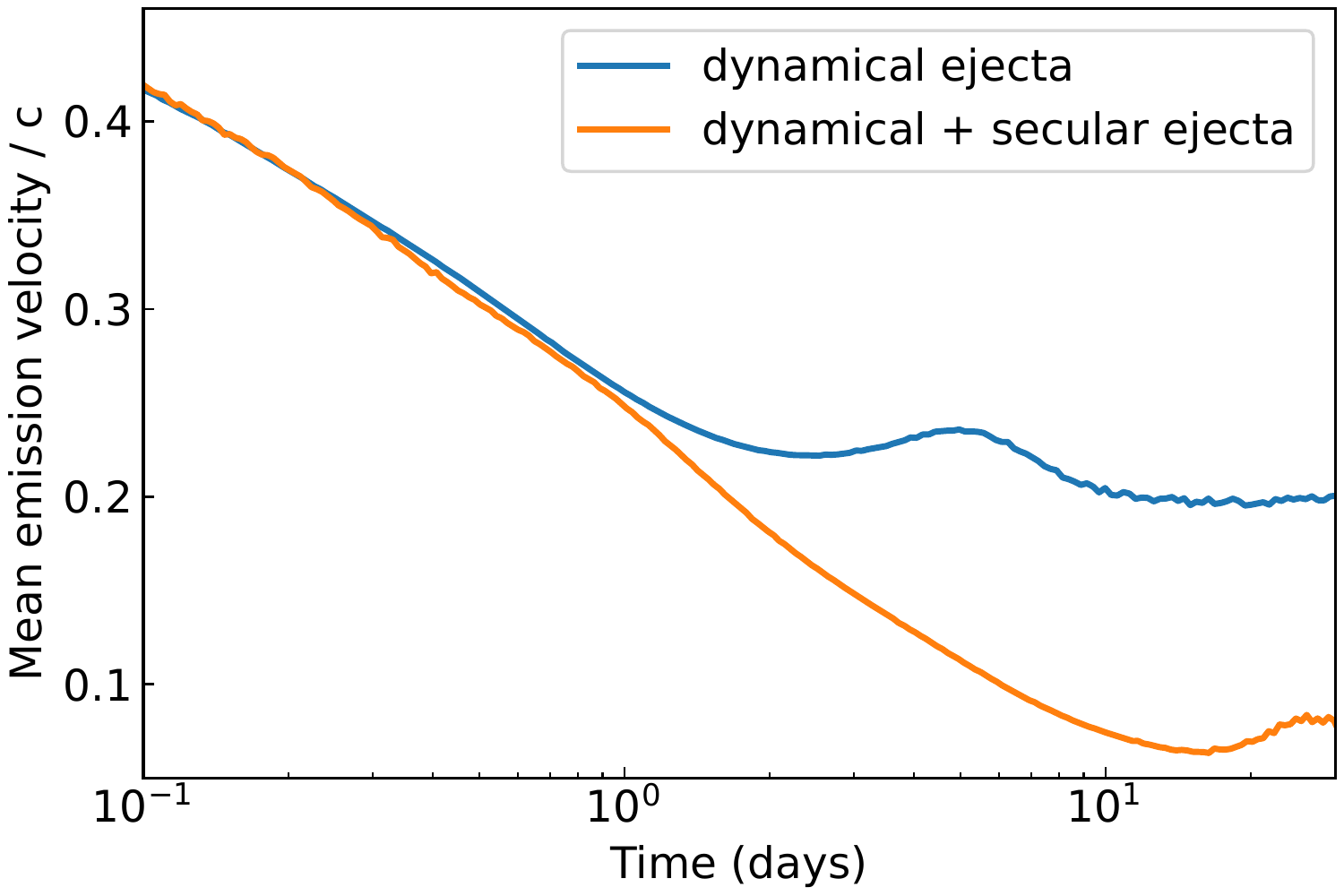}

\caption{Mean angle-averaged ejecta radial velocity ($v=rt$) of last interaction point of escaping radiation.
}

\label{fig:emission_velocity_secular_ejecta}
\end{figure}

We show the mean angle-averaged ejecta velocities at
which Monte Carlo packets are emitted from for
the simulations with and without the late time ejecta component
in Figure~\ref{fig:emission_velocity_secular_ejecta}.
At early times, the mean velocities are the same in both models,
but at later times the mean velocities
are much lower in the model with combined dynamical and late-time ejecta.
At these times the emission is primarily from the
low velocity, high density late-time ejecta component.

\subsubsection{Effect on light curve angle-dependence}

\begin{figure}
\includegraphics[width=0.5\textwidth]{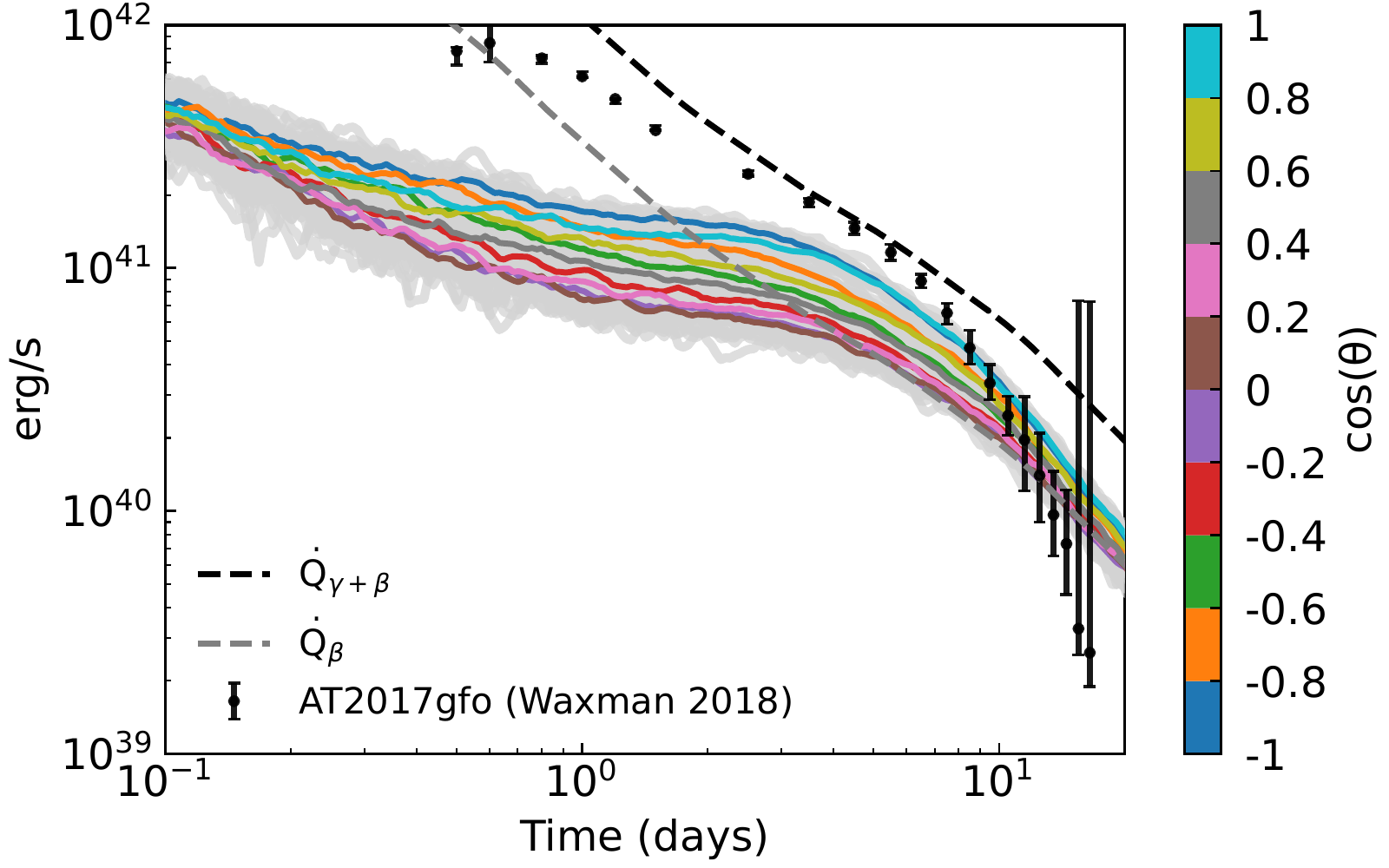}

\caption{Line of sight dependent bolometric light curves for the model including the late-time ejecta component.
{When presenting the angle-dependent light curves, we plot them as equivalent isotropic luminosities (multiplied by $4\pi \cdot$sr).}
We note that Monte Carlo noise is visible in this plot.}

\label{fig:lightcurves-angles-secularejecta}
\end{figure}

We also show the angle-dependent bolometric light curves for the model with the late-time ejecta
in Figure~\ref{fig:lightcurves-angles-secularejecta}.
We note that the higher densities, and therefore larger amount of energy at low velocities, have slightly increased
the predicted variation with observation angle.
We find that the brightest lines of sight are still in the polar directions,
which have the lowest densities and lowest opacities,
and therefore we find photons are preferentially emitted in these directions.
The faintest lines of sight are those near the equator,
as these have the
highest densities and highest opacities.
The increased viewing angle dependence is due to
more packets of photons being generated at lower velocities
in the central regions of the ejecta.
Instead of the packets escaping quickly from the lower
density dynamical ejecta
(mostly in the line of sight where the packet was created),
the packets take longer to diffuse
out of the higher density material, but preferentially in the
polar directions since these have lower opacities,
which leads to the increased angle dependence.
This highlights the need for long-term evolution simulations,
and the need to include ejecta expelled on longer timescales.
Future simulations where the late-time ejecta are self-consistently
calculated are required.

\section{Discussion and conclusions}

In this paper we aimed to predict the kilonova emission
from the ejecta density and composition directly from a neutrino-hydrodynamics simulation of a binary neutron star merger.
Initially, we considered the dynamical ejecta alone,
which are expelled up to $\sim 20$ milliseconds after the merger.
We presented bolometric light curves, and found that we do not
expect a rise to peak within the time frame of our simulation (starting at 0.02 days). This is due to the distribution
of energy generated in the ejecta.
Energy is generated and thermalised at very high velocities
and due to low optical depths is able to immediately diffuse out
of the ejecta, leading to a bolometric light curve that begins declining
from very early times after the merger.
A similar result has previously been
reported, e.g. by \citet{banerjee2020a}, \citet{klion2022a} and \citet{kawaguchi2022a}.

We do however, find a `shoulder' in the light curve,
particularly in the polar directions, around the time when
the ejecta start to
become optically thin.
\citet{just2022a} predicted light curves for a similar model
to the dynamical ejecta model we consider here (they also consider a 1.35-1.35 M$_\sun$ model using the SFHo equation of state and the neutrino leakage scheme ILEAS).
They also find a `shoulder' feature at a similar time, although their shoulder luminosity is lower than ours by a factor of a few.

We find that light curves are brighter in the polar directions
around the time of the shoulder compared to the equatorial 
lines of sight.
After the shoulder, once the ejecta become optically thin
we no longer see any angle dependence.
The angle variation is similar to the level found by \citet{just2022a} and \cite{kawaguchi2021a}.

We assumed that all energy came from $\beta$-decays.
We found that \mbox{$\gamma$-rays} are only thermalised within approximately the first hour after the merger.
The late time light curve (after the shoulder) closely follows the injected heating rate, since the ejecta have become optically thin. 
In future simulations it will be important
to calculate the fraction of thermalised heating energy self-consistently.

We showed that in the day after the merger, most Monte Carlo packets of radiation were emitted
from the outer layers of the ejecta, which at early times resembles a photosphere.
This suggests that photospheric approximations may be reasonable
at early times.
We note that the angular dependence of this photospheric-like emission
(e.g. 0.5 days)
does not necessarily lead to strong asymmetries in the light curves
(i.e. the structure of the photospheric-like emission does not correspond to structure in the angle variation of the light curves). 
Due to the high expansion velocities, the density of the ejecta
quickly decreases, as does the optical depth, leading to lower
emission velocities.
By $\sim$ 2 days we find packets are emitted from all velocities in the
ejecta, and the emission no longer resembles a photosphere.

The ejecta temperatures also rapidly decrease,
again due to the high expansion velocities.
Although we do not carry out frequency dependent simulations,
we estimate the frequency of escaping packets from the radiation
temperature in the cell from which the packet escaped. 
We did this by sampling a black body distribution at the radiation 
temperature.
We generated band light curves based on this, and showed that
due to the temperature evolution we find a rapid evolution from
blue to red colours, which is similar to that observed in AT2017gfo,
however, the evolution is on much shorter timescales.
This suggests that cooling of the ejecta could drive the colour evolution.
However, frequency dependent simulations of models including 
self-consistently evolved late time ejecta will be required 
to investigate this further.
Since this model was not tuned in any way to match the observations of AT2017gfo, it is promising that we find a similar colour evolution from our model with relatively simple assumptions.

We also carried out a simulation where we included,
in an approximate, spherically symmetric manner, ejecta expelled
from the merger remnant.
We investigated the effect on the predicted light curves,
and the extent to which dynamical
ejecta may be responsible for the observed light curve of AT2017gfo.
We found that the light curve at early times does not show a strong
sensitivity with respect to the late-time ejecta component,
although it becomes increasingly important with time.
After $\sim$1 day the light curve is dominated by the late-time ejecta
component, due to the higher densities, and therefore higher energy in the center
of the ejecta, which is able to diffuse out at later times.
The late time light curve is of similar brightness to that of AT2017gfo,
however, we still do {not} match the initial brightness.
This suggests that higher masses of high velocity
ejecta {are required} to match the initial brightness of AT2017gfo.

We also showed that the viewing angle dependence does not disappear, and even increases, with the addition of a massive, spherically symmetric late-time ejecta. This suggests that the dynamical ejecta can have a significant impact on the angle dependence even at late times.
This highlights the importance of long-term evolution simulations
for predicting kilonovae.

\section*{Acknowledgements}
We thank Ricard Ardevol-Pulpillo and Thomas Janka for providing the ILEAS scheme. CEC, AB and OJ acknowledge support by the European Research Council (ERC) under the European Union’s Horizon 2020 research and innovation program under grant agreement No. 759253. 
AB, GMP and OJ acknowledge support by Deutsche Forschungsgemeinschaft (DFG, German Research Foundation) - Project-ID 279384907 - SFB 1245.
AB and VV acknowledge support by DFG - Project-ID 138713538 - SFB 881 (“The Milky Way System”, subproject A10). AB and GMP acknowledge support by the State of Hesse within the Cluster Project ELEMENTS.
The work of SAS was supported by the Science and Technology Facilities Council [grant numbers ST/P000312/1, ST/T000198/1].
GMP and LJS acknowledge support by the European Research Council (ERC) under the European Union’s Horizon 2020 research and innovation program (ERC Advanced Grant KILONOVA No. 885281).
This work was performed using the Cambridge Service for Data Driven Discovery (CSD3), part of which is operated by the University of Cambridge Research Computing on behalf of the STFC DiRAC HPC Facility (www.dirac.ac.uk). The DiRAC component of CSD3 was funded by BEIS capital funding via STFC capital grants ST/P002307/1 and ST/R002452/1 and STFC operations grant ST/R00689X/1. DiRAC is part of the National e-Infrastructure.
OJ is grateful for computational support by the HOKUSAI computing facility at RIKEN.
CEC, OJ and VV are grateful for computational support by the VIRGO cluster at GSI.
NumPy and SciPy
\citep{oliphant2007a}, IPython \citep{perez2007a}, Matplotlib
\citep{hunter2007a}, PyVista \citep{sullivan2019pyvista} and \href{https://github.com/artis-mcrt/artistools}{\textsc{artistools}}\footnote{\href{https://github.com/artis-mcrt/artistools/}{https://github.com/artis-mcrt/artistools/}} were used for data processing and plotting.


\section*{Data Availability}

Data will be made available upon reasonable request.




\bibliographystyle{mnras}
\bibliography{astrofritz} 




\appendix

\section{Ejecta velocity increase due to r-process heating}
\label{sec:velocityboost}

\begin{figure}
\includegraphics[width=0.5\textwidth]{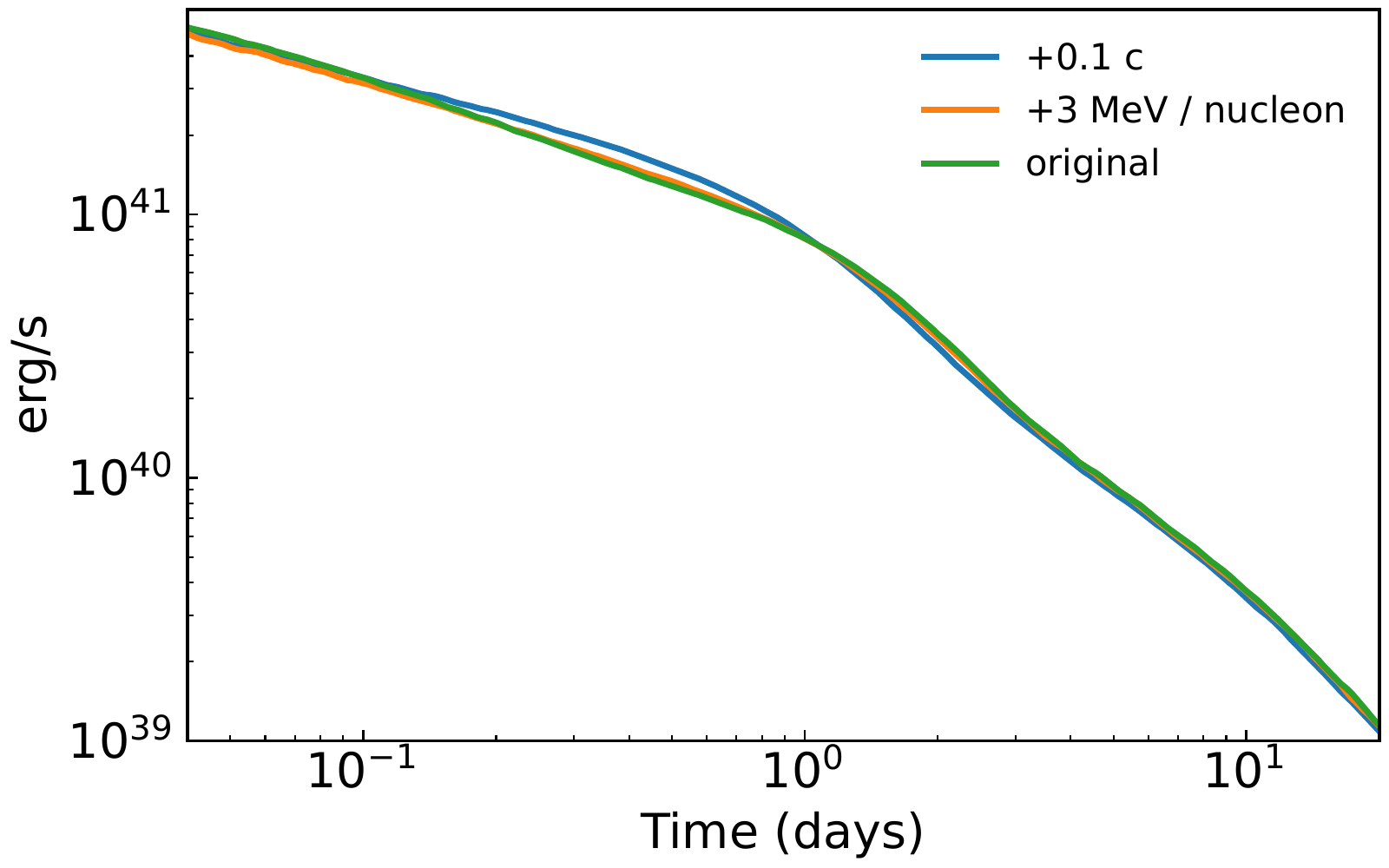}

\caption{{Angle-averaged light curves for models where we have increased SPH particle velocities to account for the potential velocity boost that could be provided by r-process heating. We show a model where we assume r-process heating can provide an additional 3 MeV per nucleon, and assume this to be kinetic energy, corresponding to an increase in SPH particle velocity of $\sim$ 0.01c. We also show a more extreme case where the velocity of each SPH particle is increased by 0.1c. These are compared to the original model with no velocity increase, described in Section~\ref{sec:model}.}}

\label{fig:lightcurves-boost}
\end{figure}

{R-process heating may provide an additional boost to the ejecta velocities, following our approximation that SPH particles are propagated after the end of the SPH simulation based on their final velocities. We have tested the potential effect of this on the light curve by assuming particles gain an additional kinetic energy of 3 MeV per nucleon from r-process heating, corresponding to an increase in velocity of $\sim$ 0.01c.  
We find that this has a negligible impact on the predicted light curve (see Figure~\ref{fig:lightcurves-boost}).
We also test a more extreme case where we assume an upper limit on the velocity boost of 0.1c (which is likely higher that r-process heating could provide).
We find that this has a small effect on the bolometric light curve.
In the angle-averaged light curves, the time at which we find the shoulder becomes $\sim 0.5$ days earlier and the brightness of the light curve at the time of the shoulder is a factor of $
\sim 2$ brighter.
Apart from at times around the shoulder, the light curves are the same.
The difference in the time and brightness of the shoulder is likely due to the lower densities resulting from the higher velocities, and therefore the optical depth is marginally lower in this case.
The qualitative results however are not affected by the increase in velocity.
}

\bsp	
\label{lastpage}
\end{document}